\documentclass[aps,pra,preprint,amsmath,amssymb,superscriptaddress,nobibnotes, longbibliography]{revtex4-1}
\usepackage{graphicx,SIunits}
\usepackage{bm}     
\usepackage{hyperref} 
\usepackage{dsfont}    
\usepackage{color}

\usepackage{amsmath}
\usepackage{amsthm}
\usepackage{amssymb}
\usepackage{tikz}

\usetikzlibrary{decorations.pathreplacing}
\usepackage{braket}
\usepackage{dsfont}

\usepackage{color}
\usepackage{lineno}
\usepackage{etoolbox}

\usepackage{multirow}

\makeatletter
\renewcommand{\email}[1]{\thanks{#1}}
\makeatother

\begin{document}


\title{Photonic variational quantum eigensolver for NISQ-compatible quantum technology}

  \author{Kang-Min Hu}%
  \email{These authors contributed equally to this work.}
      \affiliation{Center for Quantum Technology, Korea Institute of Science and Technology (KIST), Seoul 02792, Korea}
        \affiliation{Division of Quantum Information, KIST School, Korea University of Science and Technology, Seoul 02792, Korea}

  \author{Min Namkung}
  \email{These authors contributed equally to this work.}
        \affiliation{Center for Quantum Technology, Korea Institute of Science and Technology (KIST), Seoul 02792, Korea}

  \author{Hyang-Tag Lim}
      \email{hyangtag.lim@kist.re.kr}
      \affiliation{Center for Quantum Technology, Korea Institute of Science and Technology (KIST), Seoul 02792, Korea}
      \affiliation{Division of Quantum Information, KIST School, Korea University of Science and Technology, Seoul 02792, Korea}

\date{\today}

\begin{abstract}
Quantum computers have the potential to deliver speed-ups for solving certain important problems that are intractable for classical counterparts, making them a promising avenue for advancing modern computation. However, many quantum algorithms require deep quantum circuits, which are challenging to implement on current noisy devices. To address this limitation, variational quantum algorithms (VQAs) have been actively developed, enabling practical quantum computing in the noisy intermediate-scale quantum (NISQ) era. Among them, the variational quantum eigensolver (VQE) stands out as a leading approach for solving problems in quantum chemistry, many-body physics, and even integer factorization. The VQE algorithm can be implemented on various quantum hardware platforms, including photonic systems, quantum dots, trapped ions, neutral atoms, and superconducting circuits. In particular, photonic platforms offer several advantages: they operate at room temperature, exhibit low decoherence, and support multiple degrees of freedom, making them suitable for scalable, high-dimensional quantum computation. Here we present methodologies for realizing VQE on photonic systems, highlighting their potential for practical quantum computing. We first provide a theoretical overview of the VQE framework, focusing on the procedure for variationally estimating ground state energies. We then explore how photonic systems can implement these processes, showing that a wide variety of problems can be addressed using either multiple qubit states or a single qudit state.
\end{abstract}


\maketitle

\section{Introduction}\label{sec1}

Quantum computers have emerged as a promising platform for advancing modern computational technologies. They offer dramatic speed-ups over classical computers for certain problems--for example, factoring large integers exponentially faster~\cite{shor1994,shor1997}, or searching an unsorted database with quadratic speed-up~\cite{grover1996,bennett1997,ambainis2004}. As envisioned by Feynman~\cite{feynman1982} and {Lloyd}~\cite{lloyd1996}, quantum systems are naturally suited for efficiently simulating other quantum systems. In this context, quantum phase estimation (QPE) {that evaluates the eigenvalues of a Hamiltonian exponential, thereby retrieving those of the Hamiltonian,} has been established as a powerful algorithm for quantum simulation~\cite{kitaev1995,abrams1999,aspuru-guzik2005,nielson2009}. 

In parallel with theoretical developments, significant experimental progress has been made across multiple hardware platforms, including trapped ions~\cite{bruzewicz2019,cirac1995}, superconducting circuits~\cite{krantz2019}, photonic systems~\cite{obrien2007,kok2007,barz2015,luo2023}, and quantum dots~\cite{loss1998}. Among them, photonic quantum computing holds unique advantages: it allows for room-temperature operation, straightforward implementation of quantum gates via linear optics~\cite{knill2001,kok2007}, supports multiple degrees of freedom (enabling hyperentanglement)~\cite{achatz2023}{. In particular, the photonic systems facilitate flying qubits, and thus advantageous for network-based quantum information and computation tasks~\cite{kimble2008,wehner2018}, including distributed quantum sensing~\cite{kim2024b,kim2025} and distributed quantum computing~\cite{divincenzo1995,ladd2010,luo2024}. This allows multiple small quantum devices to work together to carry out tasks that normally require a large quantum system, with the additional advantage of quantum-applied security \cite{huang2017}.}

Despite this progress, many quantum algorithms require deep circuits, which remain impractical under noisy conditions on current hardware. While quantum error correction (QEC) is a long-term solution~\cite{nielson2009,lidar2013,steane1996,raussendorf2006,raussendorf2007,fowler2009}, its resource overhead is substantial, especially for photonic systems where generating high-fidelity entanglement remains challenging~\cite{gottesman2001,cochrane1999,leghtas2013}. This has motivated the development of noisy intermediate-scale quantum (NISQ) algorithms~\cite{preskill2018}, which are designed to work with shallow circuits and limited qubit counts. Variational quantum algorithms (VQAs)~\cite{cerezo2021} are a prominent class of such algorithms, combining quantum circuits with classical optimization loops to find solutions. For instance, the quantum approximate optimization algorithm can efficiently solve combinatorial optimization problems~\cite{farhi2014,bravyi2020,bae2024}, including traveling salesman problem that is associated with navigation system~\cite{narwadi2017}. Moreover, VQAs are applicable for identifying unknown quantum states~\cite{patterson2021,lee2023}{, analyzing a material~\cite{kang2025}, } or sensing {various quantities including qubit parameters and optical phases}~\cite{meyer2021,beckey2022,yang2022,koczor2020,kaubruegger2019,maclellan2024,cimini2024}. 

Among the various VQAs, the variational quantum eigensolver (VQE) is one of the most widely studied and practically relevant. It is particularly useful for estimating ground state energies in quantum chemistry and many-body physics. Detailed theoretical perspectives of the VQE are overviewed by several review articles~\cite{cerezo2021,tilly2022,fedorov2022,anand2022,qin2023}. Experimental demonstrations of VQE have been reported on diverse platforms, including photonic systems~\cite{peruzzo2014,wang2023,borzenkova2024,baldazzi2025,ghavami2025,santagati2018,kim2024}, quantum dots~\cite{maring2024}, trapped ions~\cite{hempel2018,zhao2023,meth2022,kawashima2021}, neutral atoms~\cite{chinnarasu2025}, and superconducting qubits~\cite{weaving2025,kandala2017,kandala2019,google2020,guo2024a,malley2016}. Photonic systems are especially well-suited for VQE due to their compatibility with room temperature operation, intrinsic scalability via high-dimensional encodings, and robustness to decoherence~\cite{baldazzi2025,kim2024}.

\begin{figure*}[t]
\centerline{\includegraphics[width=16cm]{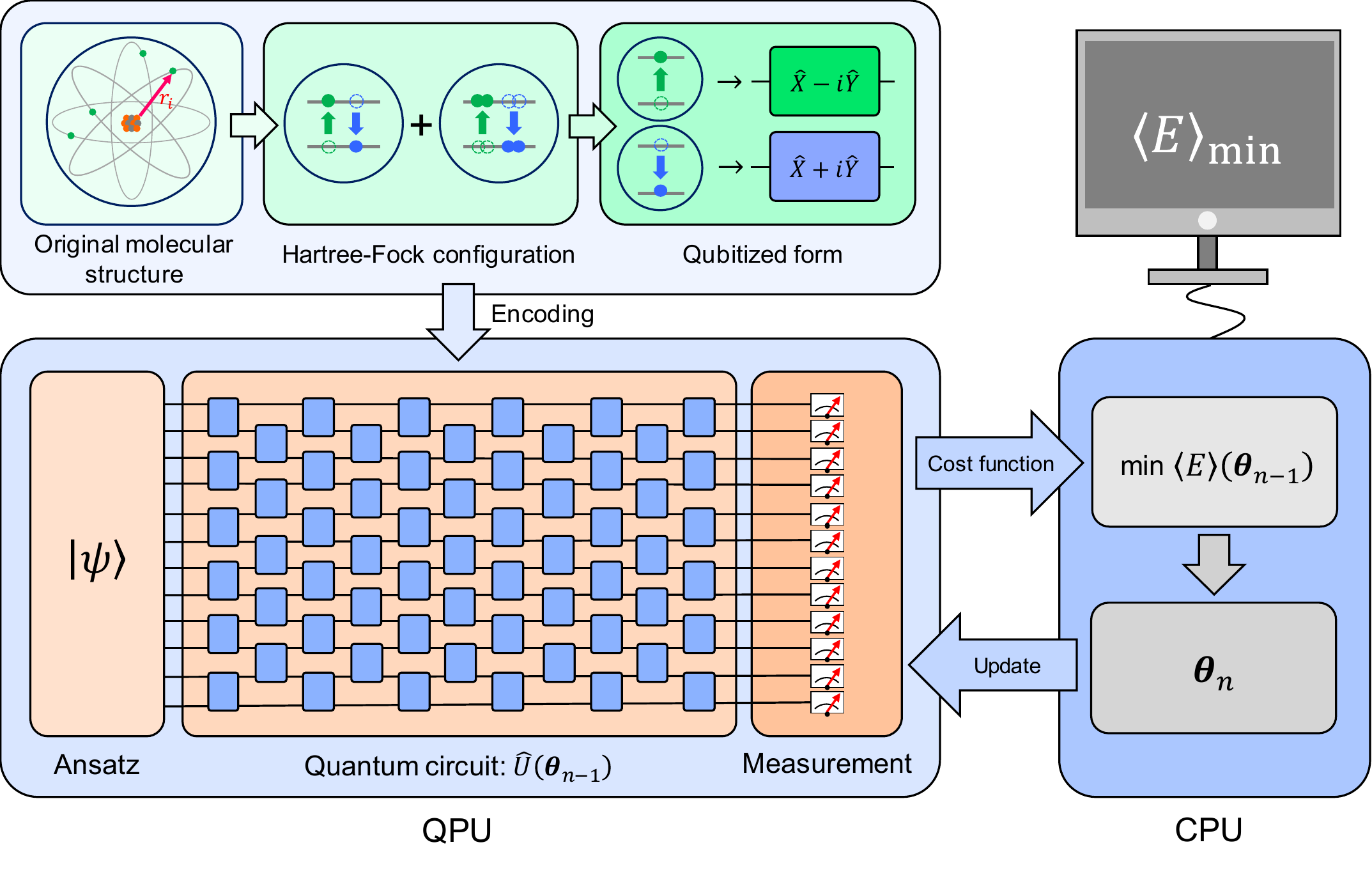}}
\caption{Overview of variational quantum eigensolver (VQE). First, in the encoding process, an original Hamiltonian for representing a molecular system, which is composed of position and momentum of a nuclei and electrons, is rewritten by Hartree-Fock configuration in terms of occupied or vacant orbitals. Then, this configuration is transformed to qubitized form, in which both annihilation and creation operators are represented in terms of $\hat{X}-i\hat{Y}$ and $\hat{X}+i\hat{Y}$ using Pauli operators $\hat{X}$ and $\hat{Y}$, respectively. The transformed structure is subsequently encoded into the quantum processing unit (QPU) parameterized by $\boldsymbol{\theta}_{n-1}$, with $n$ denoting the iteration number. After measuring the ansatz state $|\psi(\boldsymbol{\theta}_{n-1})\rangle=\hat{U}(\boldsymbol{\theta}_{n-1})|\psi\rangle$, the measurement outcome is used to update the parameter in the classical processing unit (CPU). Here, $\boldsymbol{\theta}_{n-1}$ changes to $\boldsymbol{\theta}_n$ such that the expectation value of the energy is minimized, and then $\boldsymbol{\theta}_n$ is updated to the QPU.}
\centering
\label{fig_1}
\end{figure*}

Here, we present a comprehensive overview of methodologies for implementing VQE using photonic systems. We first introduce the theoretical perspectives of the VQE, covering each process including Hamiltonian approximation, ansatz construction, Hamiltonian qubitization, Pauli measurement grouping, and numerical optimization. We also discuss error mitigation techniques--particularly zero-noise extrapolation (ZNE)~\cite{he2020,kim2021} and Pauli noise error-mitigation~\cite{urrehman2021,maciejewski2020}--that enhance the noise resilience of VQE in realistic experimental settings. Subsequently, we highlight experimental implementations of photonic VQE across different use cases, ranging from quantum chemistry and many-body systems to integer factorization. We emphasize that photonic systems not only enable efficient realization of small-scale VQE with few qubits, but also naturally support qudit-based implementations, allowing access to high-dimensional Hilbert spaces. Given these advantages, photonic VQE stands out as a NISQ-compatible quantum algorithm and a promising approach toward practical quantum computing.

\section{Theoretical background}\label{sec2} 

\subsection{Objective and basic principle of VQE}
In this section, we briefly review the theoretical perspectives of the VQE, as schematically illustrated in Fig.~\ref{fig_1} and discussed in previous review articles~\cite{mcardle2020,cerezo2021,tilly2022,fedorov2022,anand2022}. The primary objective of the VQE is to estimate the ground state energy of a given quantum system. This is achieved by variationally controlling a trial wave function $|\psi(\boldsymbol{\theta})\rangle$ to minimize the expectation value of the system's Hamiltonian:
\begin{equation}\label{e}
    E(\boldsymbol{\theta}) =\frac{\langle\psi(\boldsymbol{\theta})| \hat{H}|\psi(\boldsymbol{\theta})\rangle}{\langle\psi(\boldsymbol{\theta})|\psi(\boldsymbol{\theta})\rangle}.
\end{equation}
Here, $\hat{H}$ is the Hamiltonian of the system. According to the variational principle, $E(\boldsymbol{\theta})$ is always greater than or equal to the true ground state energy, and thus minimizing $E(\boldsymbol{\theta})$ over $\boldsymbol{\theta}$ leads to the ground state energy. Compared to QPE {generating superposition for encoding all eigenvalue exponentials via controlled unitary operations}~\cite{kitaev1995,abrams1999,nielson2009}, which requires deep quantum circuits with {$O(\epsilon^{-1})$ quantified in terms of the circuit-depth complexity $O$} to achieve accuracy $\epsilon$, the VQE is NISQ-compatible. It requires only constant-depth quantum circuits $O(1)$, at the cost of repeated measurements $O(\epsilon^{-2})$~\cite{tilly2022}, making it more practical for noisy quantum hardware.

\subsection{Hamiltonian approximation}
VQE is applicable to a variety of quantum systems, including many-body models (e.g., Heisenberg models)~\cite{kandala2017,jattana2022,lyu2023} and molecular Hamiltonians of interest in quantum chemistry~{\cite{tilly2022}}. To simulate a molecular system, it is necessary to transform its fermionic Hamiltonian into a qubit-representable form through several approximations.

First, we apply the Born--Oppenheimer approximation~\cite{born1927} to fix the nuclei and simplify the system into its electronic Hamiltonian:
\begin{eqnarray}\label{h}
 \hat{H}=\underbrace{-\frac{\hbar^2}{2m_i^2}\sum_{i}\nabla_i^2}_{\hat{T}_{\rm e}} \   \underbrace{-\frac{1}{4\pi\epsilon_0}\sum_{j,k}\frac{e^2Z_k}{|\boldsymbol{r}_j-\boldsymbol{{R}_k}|}}_{\hat{V}_{\rm ne}} \ {\underbrace{+\frac{1}{8\pi\epsilon_0}\sum_{j\not=k}\frac{e^2}{|\boldsymbol{r}_j-\boldsymbol{r}_k|}}_{\hat{V}_{\rm ee}}},  
\end{eqnarray}
with the Planck constant $\hbar$, the permittivity of free space $\epsilon_0$, position and mass of an $i$-th electron $\boldsymbol{r}_i$ and $m_i$, an atomic number of $k$-th nuclei $Z_k$, an elementary charge $e$, and a Laplace operator of the $i$-th electron $\nabla_i^2$. Let us denote the $j$-th spin-orbital of the system as $\phi_j(\boldsymbol{r}_j)\in\{0,1\}$, where $\phi_j=1$ ($\phi_j=0$) means that the orbital is occupied (unoccupied) by an electron. In the first quantization, the wave function is represented by $|\psi\rangle=|\phi_1\phi_2\cdots\phi_n\rangle$ in a Hilbert space~\cite{tilly2022,mcardle2020}, and the Hamiltonian of Eq.~(\ref{h}) is rewritten as
\begin{eqnarray}\label{h_1st}
    \hat{H}=\sum_{i=1}^{m}\sum_{p,q=1}^{n}h_{pq}|\phi_p^{(i)}\rangle\langle\phi_q^{(i)}|+\frac{1}{2}\sum_{j\not=k}^{m}\sum_{p,q,r,s=1}^{n}h_{pqrs} |\phi_p^{(i)}\phi_{q}^{(j)}\rangle\langle\phi_r^{(i)}\phi_{s}^{(j)}|,
\end{eqnarray}
where $h_{pq}=\langle\phi_p| \hat{T}_{\rm e}+\hat{V}_{\rm ne}|\phi_q\rangle$ and $h_{pqrs}=\langle\phi_{p}\phi_q|\hat{V}_{\rm ee}|\phi_r\phi_s\rangle$. Although the above Hamiltonian is represented in terms of qubit basis vectors $|\phi_j\rangle$ ($\phi_j=0,1$), a quantum circuit having deep circuit depth is required to estimate the ground state energy~\cite{tilly2022}. For this reason, we need to consider the second quantization, eventually leading to the Hamiltonian whose ground state energy is efficiently estimated with quantum circuit having shallow circuit depth. In the second quantization, both the occupation ($\phi_j=1$) and the vacancy ($\phi_j=0$) are represented in terms of Fermion creation and annihilation operators $\hat{a}_j^\dagger$ and $\hat{a}_j$, respectively. It guides the Hamiltonian of Eq.~(\ref{h_1st}) to the Hartree-Fock approximation:
\begin{equation}\label{h_2nd}
    \hat{H}=\sum_{p,q}h_{pq}\hat{a}_p^\dagger\hat{a}_q+\frac{1}{2}\sum_{p,q,r,s}h_{pqrs}\hat{a}_p^\dagger\hat{a}_q^\dagger\hat{a}_r\hat{a}_s.
\end{equation}
{Remarkably, the second quantization pursuits the operator representation instead of wavefunctions, meaning that spatial integration is not needed~\cite{moll2018}. It also implies that several issues such as antisymmetry is solved well using the second quantization~\cite{tilly2022}, rather than the first one, eventually enabling the efficient circuit-depth complexity.}

\subsection{Design of variational ansatz}
A central component of the VQE is the construction of a trial wave function (ansatz) $|\psi\rangle$, which is optimized to minimize the energy.  One widely adopted approach is the hardware-efficient ansatz (HEA)~\cite{jattana2022,leone2024}, which builds the ansatz using parameterized quantum circuit $\hat{U}(\boldsymbol{\theta})$ compatible with the available quantum hardware, enabling shallow circuit depth. Hartree-Fock method is also useful to describe the ansatz in terms of the occupation and the vacancy of electrons as in Eq.~(\ref{h_2nd}). In full configuration interaction (FCI), the ansatz is described as superposition {among} all possible electron configurations~\cite{tilly2022,mcardle2020}:
\begin{eqnarray}\label{fci}
    |\psi\rangle=c_0|\psi\rangle_{\rm ref}+\sum_{i,a}c_{ia}\hat{a}_a^\dagger \hat{a}_i|\psi\rangle_{\rm ref}+\sum_{i,j,a,b}c_{ijab}\hat{a}_a^\dagger\hat{a}_b^\dagger\hat{a}_i\hat{a}_j|\psi\rangle_{\rm ref}+\cdots,
\end{eqnarray}
where $|\psi\rangle_{\rm ref}$ denotes the reference Hartree-Fock configuration. Here, the second and the third summations are associated with singly-excited and doubly-excited configurations, respectively. According to the coupled cluster  theory~\cite{bartlett2007,coester1960,cizek1966,paldus1972,paldus1977,cizek1980,paldus1984}, the FCI represented in Eq.~(\ref{fci}) is simplified as 
\begin{equation}\label{cc}
    |\psi\rangle=e^{\hat{T}}|\psi\rangle_{\rm ref},
\end{equation}
where $\hat{T}$ is a cluster (or truncated) operator
\begin{eqnarray}
    \hat{T}&=&\sum_{n}\hat{T}_n \ \textrm{with} \nonumber\\
    \hat{T}_1&=&\sum_{i,a}t_{ia}\hat{a}_i^\dagger\hat{a}_a, \ \hat{T}_2=\sum_{i,j,a,b}t_{ij,ab}\hat{a}_i^\dagger\hat{a}_j^\dagger\hat{a}_a\hat{a}_b, \ \cdots.
\end{eqnarray}
By using the FCI, it is possible to accurately estimate the ground state energy. However, the FCI in Eqs.~(\ref{fci}) and (\ref{cc}) contains a lot of complex multiply-excited configurations, making the estimation be very inefficient. To improve the efficiency, it is assumed that both terms corresponding to single and double-excitations are more dominant than high-order terms, leading to coupled cluster single and double (CCSD) structure: 
\begin{equation}
    \hat{T}_{\rm CCSD}=\hat{T}_1+\hat{T}_2.
\end{equation}
We note that an exponential operator in Eq.~(\ref{cc}) is not a unitary operator, implying that it is not realized using a quantum circuit. To circumvent this, one can establish unitary coupled cluster single and double (UCCSD) structure as~\cite{bartlett1989,taube2006,wecker2015,lee2018,filip2020}
\begin{equation}
    |\psi\rangle=e^{\hat{T}-\hat{T}^\dagger}|\psi\rangle_{\rm ref}.
\end{equation}

{While the UCCSD ansatz provides a chemically motivated and systematically improvable framework that has become a standard choice for quantum simulations of molecular systems, its practical use on NISQ hardware can be limited by the large number of parameters, the associated circuit depth, and the need for Trotterization. In addition, as a single-reference ansatz truncated at single and double excitations, UCCSD may be less effective for strongly correlated systems, where more compact, generalized, or adaptive ansatz can offer improved accuracy and resource efficiency~\cite{romero2019,grimsley2020}.}

\subsection{Qubitization of the Hamiltonian}
To simulate an electronic system using quantum circuits, the fermionic Hamiltonian must be mapped onto a qubit Hamiltonian defined over the Hilbert space $(\mathbb{C}^2)^{\times d}$. This process is called Hamiltonian qubitization. A widely used transformation is the Jordan--Wigner transformation~\cite{jordan1928}, which maps fermion operators $\hat{a}_j^\dagger$ and $\hat{a}_j$ as:
\begin{eqnarray}
    \hat{a}_j^\dagger&\rightarrow&\frac{\hat{X}_j-i\hat{Y}_j}{2}\otimes \hat{Z}_0\otimes\cdots\otimes\hat{Z}_{j-1}, \nonumber\\
    \hat{a}_j&\rightarrow&\frac{\hat{X}_j+i\hat{Y}_j}{2}\otimes \hat{Z}_0\otimes\cdots\otimes\hat{Z}_{j-1}.
\end{eqnarray}
This yields the qubitized Hamiltonian as a linear combination of Pauli strings $\hat{P}_a$ on $(\mathbb{C}^{2})^{\times d}$:

\begin{equation}
    \hat{H}=\sum_{a}w_a \hat{P}_a,
\end{equation}
with {real} coefficients $w_a$. 

\begin{figure*}[t]
\centerline{\includegraphics[width=16cm]{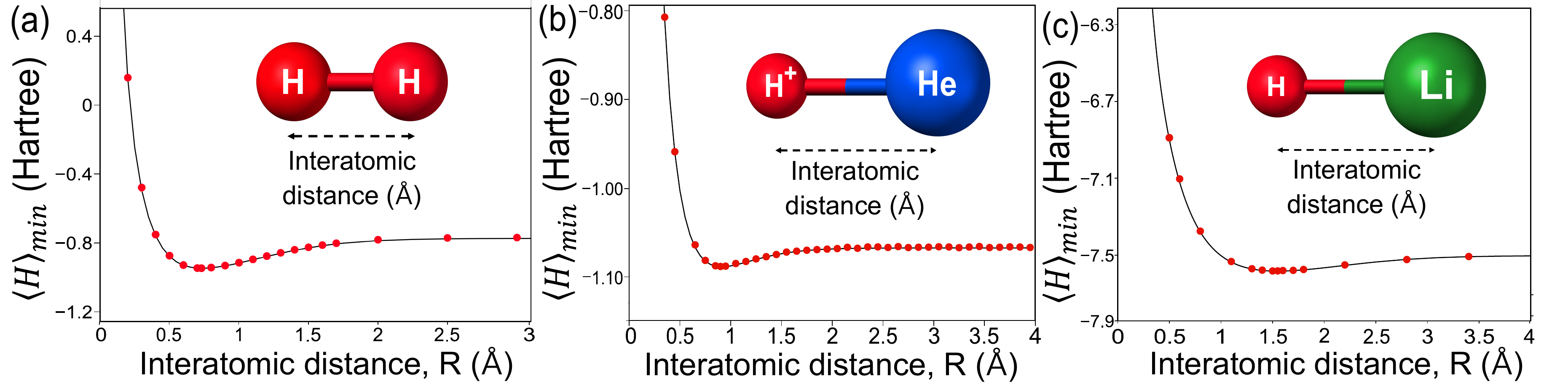}}
\caption{Simulated results of quantum chemistry problems. Red points are energy values estimated by the VQE, and black curved lines show the theoretical values of ground energies. (a) $\mathrm{H}_2$ model. (b) ${\rm HeH^+}$ model. (c) LiH model.}
\centering
\label{fig_2}
\end{figure*}

For instance, the ground energies of $\mathrm{H}_2$ and $\mathrm{He}\mathrm{H}^+$ are efficiently estimated from the simple Hamiltonians~\cite{maring2024,skryabin2023,kim2024}
\begin{eqnarray}
    \hat{H}_{\rm H_2}&=&w_0 \hat{I}\otimes\hat{I}+w_1 \hat{I}\otimes\hat{Z}+w_2\hat{Z}\otimes\hat{I}\nonumber\\
    &+&w_3\hat{Z}\otimes\hat{Z}+w_4\hat{X}\otimes\hat{X},
\end{eqnarray}
and
\begin{eqnarray}    
    \hat{H}_{\rm HeH^+}&=&w_0 \hat{I}\otimes\hat{I}+w_1 \hat{I}\otimes\hat{Z}+w_2\hat{Z}\otimes\hat{I}\nonumber\\
    &+&w_3\hat{Z}\otimes\hat{Z}+w_4\hat{X}\otimes\hat{X}+w_5\hat{I}\otimes\hat{X} \nonumber\\
    &+&w_6\hat{Z}\otimes\hat{X}+w_7\hat{X}\otimes\hat{I}+w_8\hat{X}\otimes\hat{Z}.
\end{eqnarray}
These expressions are typically derived using Jordan-Wigner or Bravyi-Kitaev transformations~\cite{seeley2012}. Figure~\ref{fig_2} illustrates simulations of VQE forestimating  ground state energies of $\mathrm{H}_2$, $\mathrm{HeH}^+$, and $\mathrm{LiH}$. The red points indicate close agreement between VQE estimates and theoretical values, demonstrating its accuracy.

In the case of many-body physics, Hamiltonians often follow the Heisenberg model. For example, the Hamiltonian of two-qubit antiferromagnetic Heisenberg model is constructed as
\begin{eqnarray}
    \hat{H}_{\rm 2q}=w_1\hat{X}\otimes\hat{X}+w_2\hat{Y}\otimes\hat{Y}+w_3\hat{Z}\otimes\hat{Z}.
\end{eqnarray}
Interestingly, integer factorization can also be performed using the VQE~\cite{anschuetz2018,karamlou2021,agresti2024}. For example, factoring $35=5\times7$ can be formulated as finding the ground state energy of the following Hamiltoniann~\cite{agresti2024}: 
\begin{equation}
    \hat{H}_{5\times7}=\left[(5\times7)\hat{I}\otimes\hat{I}-\left(6\hat{I}-\hat{Z}\right)\otimes\left(6\hat{I}-\hat{Z}\right)\right]^2.
\end{equation}
{In contrast to Shor’s algorithm, whose circuit depth scales as at least $O(n^3\log n)$~\cite{takahashi2006,tan2024}, the VQE-based algorithm operates with only a constant-depth circuit~\cite{tilly2022}. This suggests that VQE makes the factorization be efficiently performed with shallow circuit depth.}

\subsection{Pauli measurement grouping}
Note that the ansatz is also transformed onto the qubit space, and thus it is generally described as $|\psi(\boldsymbol{\theta})\rangle=\hat{U}(\boldsymbol{\theta})|\boldsymbol{0}\rangle$ with $|\boldsymbol{0}\rangle=|0\rangle^{\otimes d}$ and a unitary operator $\hat{U}(\boldsymbol{\theta})$ representing the parameterized quantum circuit. It means that the expectation value of Eq.~(\ref{e}) is estimated as
\begin{eqnarray}\label{e_jw}
    E(\boldsymbol{\theta})&=&\langle\psi(\boldsymbol{\theta})|\hat{H}|\psi(\boldsymbol{\theta})\rangle\nonumber\\
    &=&\sum_{a}w_a\langle\boldsymbol{0}|\hat{U}^\dagger(\boldsymbol{\theta})\hat{P}_a\hat{U}(\boldsymbol{\theta})|\boldsymbol{0}\rangle\nonumber\\
    &=&\sum_aw_a \langle\hat{P}_a\rangle_{\boldsymbol{\theta}}.
\end{eqnarray}
Here, we consider the Abelian group consisting of certain Pauli strings to group some Pauli measurements. Specifically, for a set of generators of the Abelian group $\{\hat{P}_a\}$, there is a unitary operator that transforms the generators to a Pauli string composed of $\hat{I}$ and $\hat{Z}$. For instance, let us consider the four Pauli strings $\hat{X}\hat{X}\hat{Y}\hat{X}$, $\hat{I}\hat{I}\hat{Y}\hat{X}$, $\hat{Y}\hat{Y}\hat{Y}\hat{X}$, and $\hat{I}\hat{I}\hat{X}\hat{Y}$~\cite{tilly2022}. It can be shown that these four Pauli strings are transformed to $\hat{Z}\hat{I}\hat{I}\hat{I}$, $\hat{I}\hat{Z}\hat{I}\hat{I}$, $\hat{I}\hat{I}\hat{Z}\hat{I}$, and $\hat{I}\hat{I}\hat{I}\hat{Z}$ by a single unitary operator, respectively. It means that all these four expectation values $\langle\hat{X}\hat{X}\hat{Y}\hat{X}\rangle_{\boldsymbol{\theta}}$, $\langle\hat{I}\hat{I}\hat{Y}\hat{X}\rangle_{\boldsymbol{\theta}}$, $\langle\hat{Y}\hat{Y}\hat{Y}\hat{X}\rangle_{\boldsymbol{\theta}}$, and $\langle\hat{I}\hat{I}\hat{X}\hat{Y}\rangle_{\boldsymbol{\theta}}$ are estimated by measuring the ansatz with a single Pauli string $\hat{Z}\hat{Z}\hat{Z}\hat{Z}$.

\begin{figure*}[t]
\centerline{\includegraphics[width=16cm]{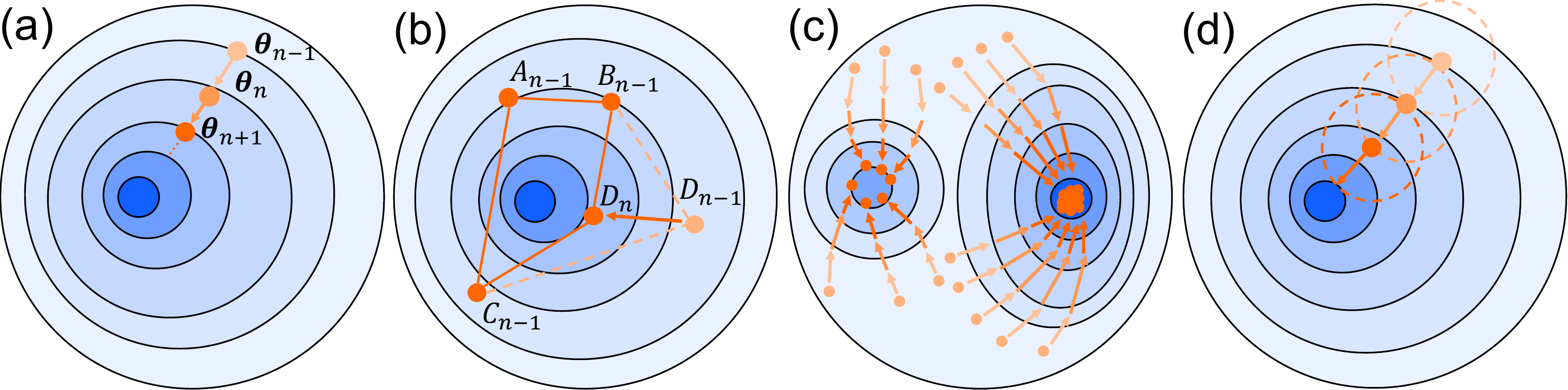}}
\caption{Description of classical optimizers. (a) Gradient-descent method, in which parameters are updated according to the iteration rule $\boldsymbol{\theta}_n=\boldsymbol{\theta}_{n-1}-\eta\vec{\nabla}E(\boldsymbol{\theta}_{n-1})$. (b) Nelder-Mead method, in which a simplex $A_{n-1}B_{n-1}C_{n-1}D_{n-1}$ iteratively contracts or expands to $A_{n-1}B_{n-1}C_{n-1}D_{n}$ according to predefined iteration rules. (c) Particle swarm optimization, in which parameters are updated simultaneously, and the ground state energy is typically estimated at the location where the largest number of parameters converge. (d) Trust region method that includes COBYLA method, in which parameters iteratively move within each trusted area.}
\centering
\label{fig_3}
\end{figure*}

\subsection{Classical optimizers for energy minimization}
The expectation value of the energy in Eq.~(\ref{e_jw}), which is estimated using a quantum computer, is numerically minimized using a classical optimizer in Fig.~\ref{fig_3}~\cite{nocedal2000}, thereby attaining the ground state energy. One well-known classical optimizer is based on the gradient-descent method~{\cite{nocedal2000}}, in which parameters $\boldsymbol{\theta}_{n-1}$ at the $n-1$-th iteration are updated to $\boldsymbol{\theta}_n$ based on the simple iteration rule: $\boldsymbol{\theta}_n=\boldsymbol{\theta}_{n-1}-\eta\vec{\nabla}E(\boldsymbol{\theta}_{n-1})$, with suitable step size $\eta>0$. Also, the Nelder-Mead method~{\cite{nocedal2000}} employs a simplex -- consisting of several parameters in the parameter space -- that contracts or expands according to predefined iteration rules, eventually converging to ground state energy without gradient evaluation. Beyond these conventional gradient-based and gradient-free methods, several improved methods are available for efficient numerical optimization. For instance, one can use simulated annealing or particle swarm optimization to improve the chance of reaching the ground state energy. Also, one can employ simultaneous perturbation stochastic approximation (SPSA)~{\cite{bhatnagar2013}}, in which stochastic approach is performed to estimate gradient from unknown parameters. Constrained optimization by linear approximation (COBYLA)~{\cite{powell2003,lee2022}} is also a promising method to efficiently estimate the ground state energy based on trust region method. {It is noted that, however, the COBYLA does not effectively circumvent the barren plateau problem, which may degrade ability of various VQAs~\cite{arrasmith2021}.}

\begin{figure}[t]
\centerline{\includegraphics[width=12cm]{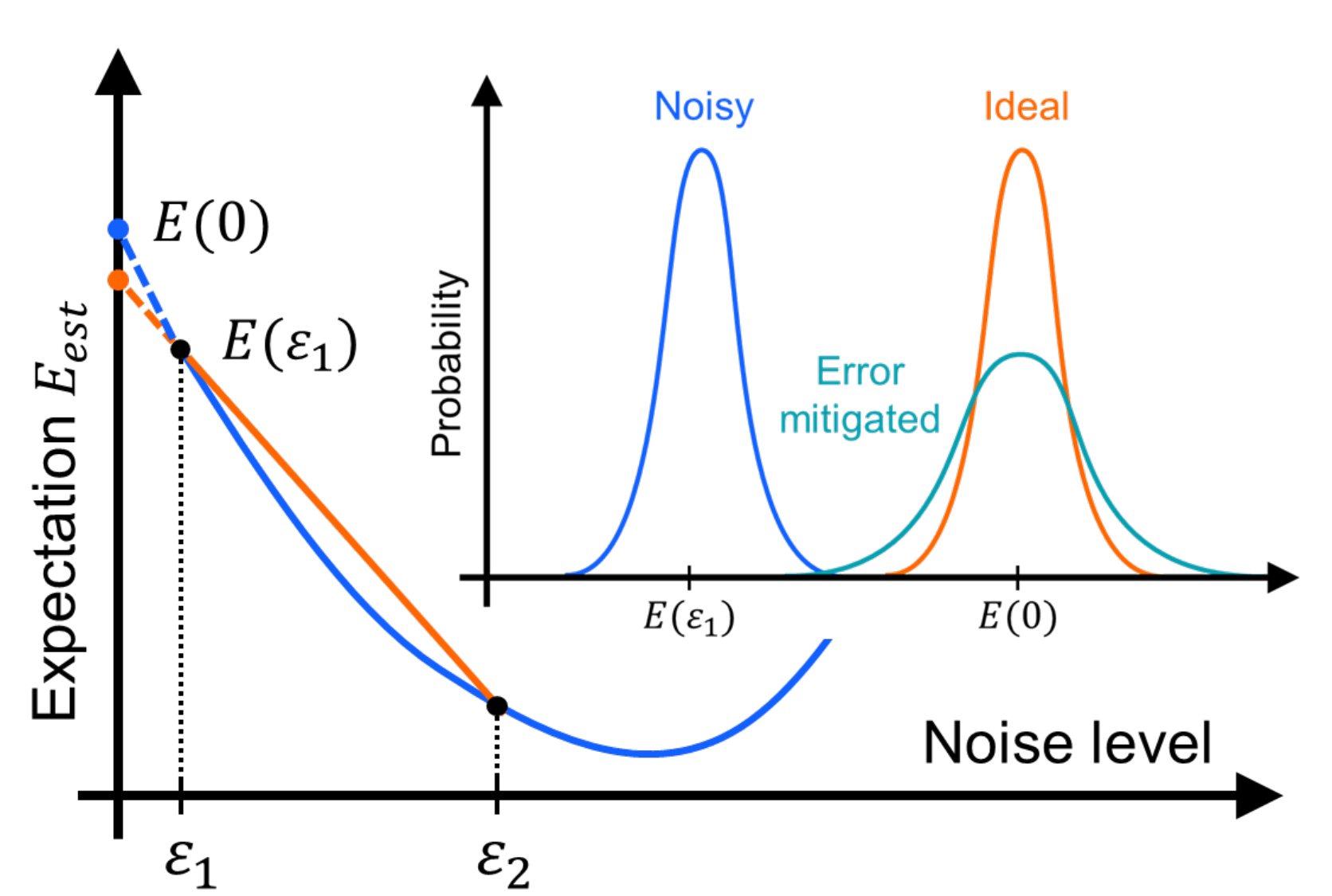}}
\caption{Concept of the zero-noise extrapolation method (ZNE), as illustrated in Ref.~\cite{borzenkova2024}. The blue curve describes the expectation value $E(\varepsilon)$ as a function of the noise strength $\varepsilon$. The orange line indicates a linear extrapolation between two measurements taken at different noise levels, $\varepsilon_1$ and $\varepsilon_2$, to estimate the zero-noise value. The inset shows the corresponding measurement probability distributions. Here, the blue, orange, and cyan curves correspond to the distributions under noise, in the ideal (noise-free) case, and the ZNE estimator $E_{\rm{est}}$, respectively.}
\centering
\label{fig_4}
\end{figure}

\subsection{Error-mitigation techniques for VQE}
As previously discussed, noise is inevitable when executing VQE on current quantum hardware. Although VQE is well-suited to the NISQ regime due to its relative robustness against noise, its performance still depends critically on the quality of state preparation, gate operations, and measurements. Therefore, integrating error-mitigation strategies is essential for enhancing the accuracy of VQE results.

One of the most widely used techniques is zero-noise extrapolation (ZNE), as illustrated in Fig.~\ref{fig_4}. This method estimates a noiseless expectation value by extrapolating measured values obtained under different noise strengths $\varepsilon$~\cite{li2017,he2020,kim2021}. For given expectation value $E(\varepsilon)$ dependent on the noise strength $\varepsilon$, the aim is to estimate the noiseless expectation value $E(0)$ based on $E(\varepsilon_1),E(\varepsilon_2),\cdots,E(\varepsilon_n)$. For example, let us assume that $E(\varepsilon)$ is approximated as
\begin{eqnarray}
    E(\varepsilon)=c_1+c_2\varepsilon.
\end{eqnarray}
Then, ZNE estimates $E_{\rm est}$ as a function written below~\cite{borzenkova2024}:
\begin{eqnarray}
    E_{\rm est}=\frac{\varepsilon_2E(\varepsilon_1)-\varepsilon_1E(\varepsilon_2)}{\varepsilon_2-\varepsilon_1},
\end{eqnarray}
which is closer to $E(0)$ than $E(\varepsilon_1)$. It is noted that as $E_{\rm est}$ approaches to $E(0)$, its variance tends to diverge. This behavior is briefly explained as follows: If we assume that both variances of $E(\varepsilon_1)$ and $E(\varepsilon_2)$ are equal to $\sigma^2$, for instance, then the variance of $E_{\rm est}$ is estimated as
\begin{eqnarray}
    \mathrm{Var}[E_{\rm est}]=\sigma^2\frac{\varepsilon_2^2+\varepsilon_1^2}{(\varepsilon_2-\varepsilon_1)^2}.
\end{eqnarray}
This means that {$\mathrm{Var}[E_{\rm est}]$ tends to diverge as $\varepsilon_1$ and $\varepsilon_2$ become close to each other during the ZNE}. {See section 3.1.3 for the details of the optical implementation.}

\begin{table}[b]
    \centering
    \begin{tabular}{ccccccc}
      Ref. & Ansatz & Dim. & Hamiltonian & Mitigation & Optimizer \\
      \hline \hline 
       \cite{peruzzo2014} & Qubits & 4 & $\mathrm{He}$-$\mathrm{H}^+$ & - & Nelder-Mead \\
       \cite{skryabin2023} & Qubits & 4 & $\mathrm{H}_2$ & - & Unknown \\
       \cite{maring2024} & Qubits & 4 & $\mathrm{H}_2$ & Pauli & Nelder-Mead \\
       \cite{borzenkova2021} & Qubits & 4 & Schwinger & - & SPSA \\
       \cite{borzenkova2024} & Qubits & 4 & Schwinger & ZNE & SPSA \\
       \cite{agresti2024} & Qubits & 4 & $5\times7$ & - & Unknown \\
       \cite{santagati2018} & Qubits & 4 & chlorophyll & - & particle-swarm \\
       \cite{ghavami2025} & Qubits & 16 & $\mathrm{He}$ & - & Unknown \\
       \cite{baldazzi2025} & Qubits & 16 & $\mathrm{H}_2$, $5\times7$ & - & gradient-descent \\
       \cite{wang2023} & Qudit & 4 & $\mathrm{He}$-$\mathrm{H}^+$ & - & SPSA-QNG \\
       \cite{lee2022} & Qudit & 4 & $\mathrm{He}$-$\mathrm{H}^+$ & Pauli & COBYLA \\
       \cite{lee2024} & Qudit & 4 & $\mathrm{He}$-$\mathrm{H}^+$ & - & COBYLA \\
       \cite{kim2024} & Qudit & 16 & $\mathrm{H}_2$, $\mathrm{LiH}$ & - & COBYLA
    \end{tabular}
    \caption{Overview of experimental results of VQE using photonic systems. In Hamiltonian column, ``$5\times7$'' denotes factorizing $35=5\times7$, ``chlorophyll'' denotes analysis of exciton transfer between chlorophyll molecules, and ``Schwinger'' denotes the Schwinger model. In mitigation column, ``Pauli'' denotes Pauli noise error-mitigation~\cite{urrehman2021}, respectively. In optimizer column, ``SPSA'' means simultaneous perturbation stochastic approximation, ``QNG'' means quantum natural gradient, and ``particle-swarm'' means particle swarm optimization.}
    \label{table:1}
\end{table}

Another method, Pauli noise error-mitigation, can be applied to error-mitigated VQE~\cite{urrehman2021,lee2022,maciejewski2020}, which is focused on recovering measurement probability distribution against measurement noise. This assumes that a noisy measurement that is experimentally implemented is described by the positive-operator-valued measure (POVM)
\begin{eqnarray}
    \hat{M}_j^{\rm (exp)}=\sum_{k}p(j|k)\hat{M}_k^{\rm (ideal)},
\end{eqnarray}
where $\hat{M}_k^{\rm (ideal)}$ are POVM elements of an ideal measurement, and $p(j|k)$ denotes the conditional probability that the correct outcome $k$ is changed to $j$ due to the measurement noise. In the above expression, $p(j|k)$ constitutes an invertible stochastic matrix
\begin{eqnarray}
    {\Lambda=\begin{bmatrix}
        p(1|1) & p(1|2) & \cdots & p(1|m) \\
        p(2|1) & p(2|2) & \cdots & p(2|m) \\
        \vdots & \vdots & \ddots & \vdots \\
        p(m|1) & p(m|2) & \cdots & p(m|m)
    \end{bmatrix}.}
\end{eqnarray}
According to Born's rule, the measurement probability distribution is described by $\boldsymbol{p}^{\rm (exp)}=(\langle\psi|\hat{M}_1|\psi\rangle,\cdots,\langle\psi|\hat{M}_m|\psi\rangle)$. Using the stochastic matrix $\Lambda$, $\boldsymbol{p}^{\rm (exp)}$ is rewritten as
$\boldsymbol{p}^{\rm (exp)}=\Lambda\boldsymbol{p}^{\rm (ideal)}.$ Therefore, by the inverse of the stochastic matrix $\Lambda^{-1}$, the ideal measurement probability distribution is fully recovered as $\boldsymbol{p}^{\rm (ideal)}=\Lambda^{-1}\boldsymbol{p}^{\rm (exp)}$. {See section 3.2.2 for the details of the optical implementation.}

\section{Photonic VQE}\label{sec3}

\subsection{{Applications}}
\subsubsection{Proposal and demonstration of VQE}
We now focus on experimental methodologies for implementing the VQE using photonic systems, as summarized in Table~\ref{table:1}. The VQE was first proposed and experimentally demonstrated in 2014 using a photonic integrated circuit to estimate the ground state energy of the $\mathrm{HeH^+}$ molecule~\cite{peruzzo2014}. 

\begin{figure*}[b]
\centerline{\includegraphics[width=15cm]{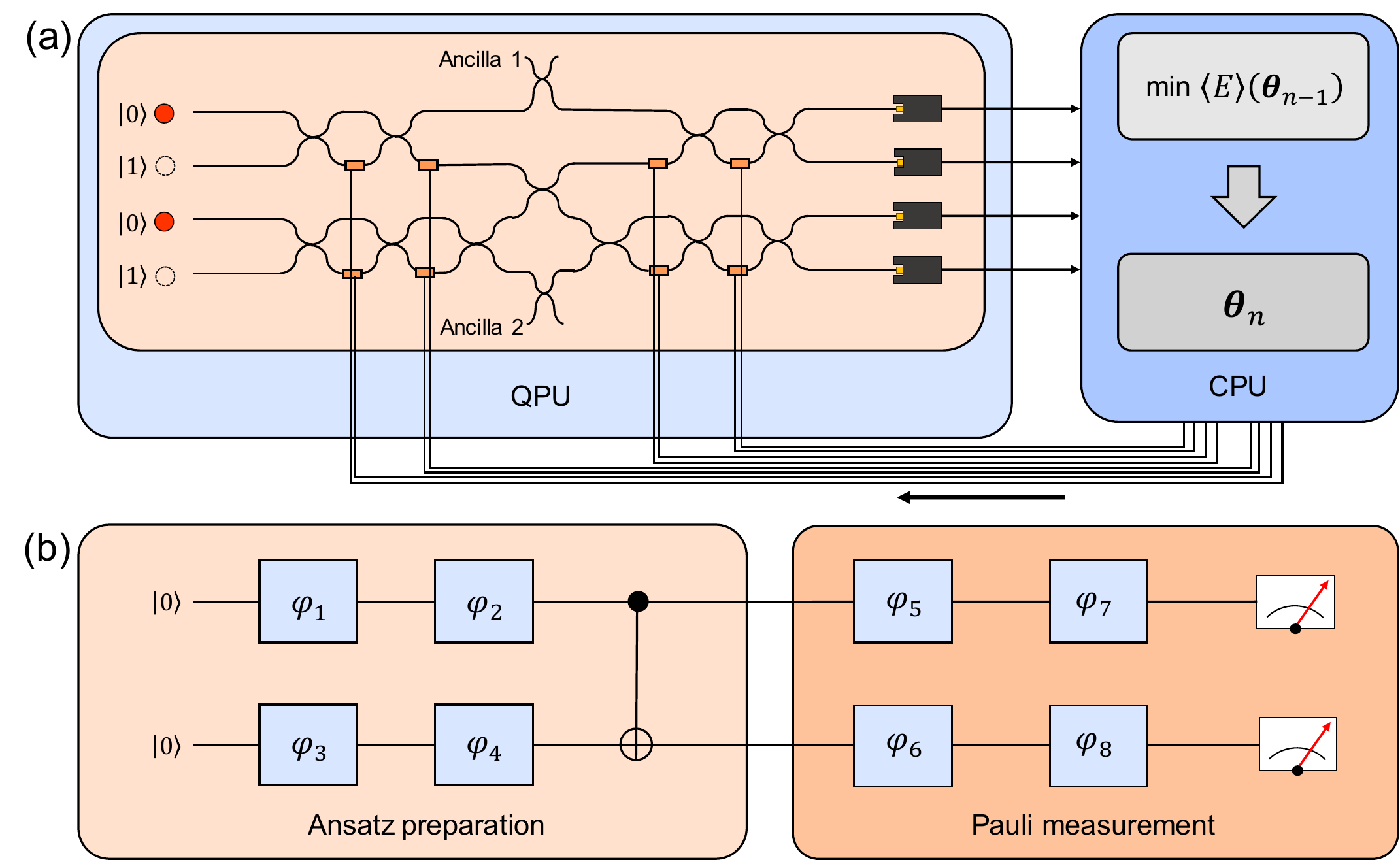}}
\caption{(a) Experimental scheme for implementing the VQE to estimate the ground state energy of $\mathrm{HeH^+}$ using two path qubits on a photonic integrated circuit~\cite{peruzzo2014,skryabin2023}. In the QPU, initial two-qubit states $|00\rangle$, generated via spontaneous parametric down-conversion (SPDC), are modulated by phase shifters controlled by a CPU. A beam-splitter network is used for entangling these qubits. Finally, the two-qubit state entangled by the beam-splitter network is measured by photodetectors that extract the which-way information, thereby estimating expectation value $\langle\psi|\hat{A}\otimes\hat{B}|\psi\rangle$ with $\hat{A},\hat{B}\in\{\hat{X},\hat{Y},\hat{Z}\}$. (b) Quantum circuit implemented by the photonic integrated circuit in (a). The blue boxes represent single-qubit gates controlled by phases $\varphi_j$.}
\centering
\label{fig_5}
\end{figure*}

In this demonstration, an ansatz was prepared using two path qubits, as illustrated in Fig.~\ref{fig_5}(a). The QPU initialize two-qubit states $|00\rangle$, generated via spontaneous parametric down-conversion (SPDC), and these were modulated by phase shifters controlled by a CPU. The expectation value was variationally minimized using the Nelder--Mead optimization method~{\cite{nocedal2000}}. Unlike gradient-descent algorithms, Nelder--Mead is well-suited for unconstrained minimization even in the presence of noise and non-smooth objective functions, making it suitable for early VQE experiments under imperfect conditions. This setup implemented the quantum circuit shown in Fig.~\ref{fig_5}(b)~\cite{peruzzo2014}. In this configuration, the path qubits are entangled via a beam-splitter network, following the design in Ref.~\cite{obrien2009}. The entanglement was achieved using only linear optical components with post-selection, a technique known as the Knill-Laflamme-Milburn (KLM) scheme~\cite{knill2001}.

Notably, the VQE is particularly effective in quantum chemistry problems even with only a few qubits. Recent experiments have estimated the ground state energy of $\mathrm{H}_2$ using similar photonic integrated circuits~\cite{maring2024,skryabin2023}. In Ref.~\cite{skryabin2023}, the fabricated circuit performed a quantum transformation whose output was analyzed via the Sinkhorn--Knopp algorithm, yielding a fidelity of $99.18\%$ with respect to the ideal mode transformation. In another study~\cite{maring2024}, an integrated photonic chip estimated the ground state energy of $\mathrm{H}_2$ with an accuracy of $\pm0.00158$ Ha, using 400,000 post-processed two-photon repetitions. The Nelder--Mead algorithm was again employed for optimization. 




\subsubsection{VQE with another quantum algorithm}

\begin{figure*}[t]
\centerline{\includegraphics[width=16cm]{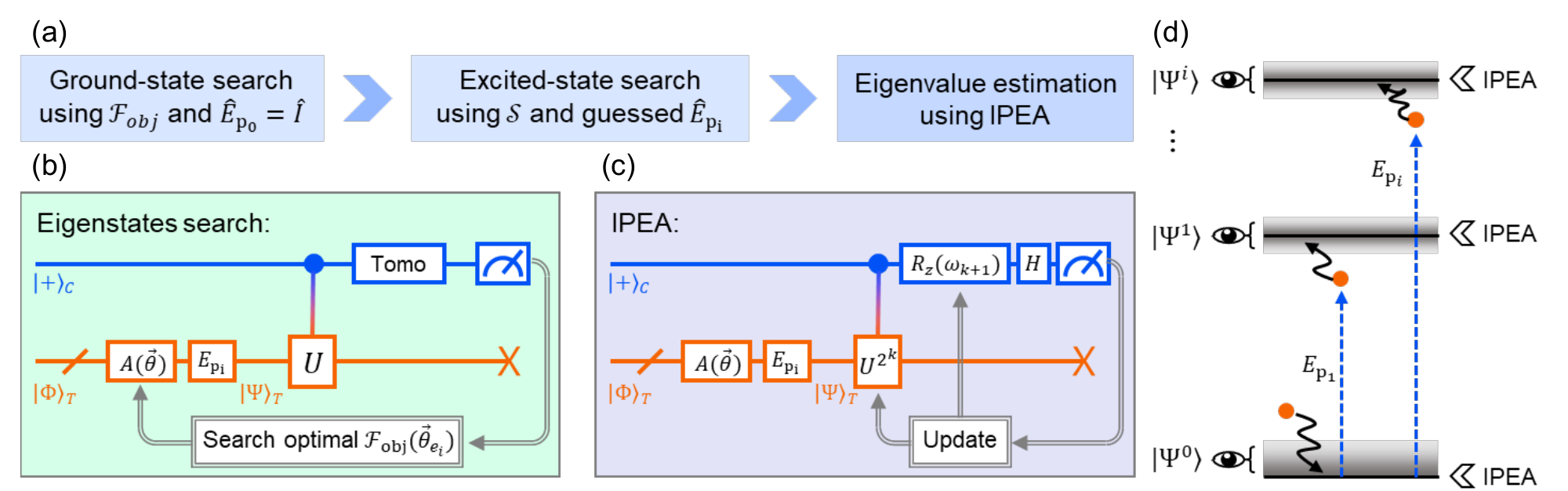}}
\caption{(a) Flowchart for WAVES algorithm in Ref.~\cite{santagati2018}. (b) Variational search for estimating the ground state energy level. (c) Iterative phase estimation algorithm (IPEA) for searching the excited energies. (d) Diagram for the WAVES approach. The initial guesses of the states are refined using the witness and IPEA returns the eigenvalues.}
\centering
\label{fig_6}
\end{figure*}

Ref.~\cite{santagati2018} introduced the Witness-Assisted Variational Eigenspectra Solver (WAVES), a hybrid quantum-classical algorithm designed to estimate both ground and excited states of a given Hamiltonian. The protocol proceeds in the following steps:

\begin{enumerate}
    \item \textbf{Ground state search}: A variational optimization is performed with an ansatz $|\psi(\boldsymbol{\theta})\rangle = \hat{U}(\boldsymbol{\theta})|\phi\rangle$, where $\hat{U}(\boldsymbol{\theta})$ is a parameterized unitary operator and $|\phi\rangle$ is a reference state. The optimization aims to minimize the free-energy-inspired cost function $F_{\mathrm{obj}}(\mathcal{P}, \varepsilon) = \varepsilon - T \mathcal{P}$, where $\varepsilon$ is an energy estimator estimated from the off-diagonal elements of the reduced density matrix of the control qubit, $\mathcal{P} = \mathrm{Tr}[\hat{\rho}_C^2]$ is the purity of the control qubit $\hat{\rho}_C$ after the controlled evolution, and $T$ is a tunable parameter that balances energy minimization and entropy suppression.
    In the ideal case, the optimal state has unit purity and yields a variationally accessible eigenstate of Hamiltonian.

    \item \textbf{Excited state search}: A unitary operator for an approximate $i$-th target excited  state is constructed via $\hat{E}_{p_i}\hat{U}(\boldsymbol{\theta}_g)$, with $\hat{E}_{p_i}$ corresponding to a system-dependent perturbation. The parameter $\boldsymbol{\theta}_{e_i}$ that minimizes $F_{\mathrm{obj}}$ in the high-$T$ limit is searched with variational approach, obtaining the unitary for the target excited  state $\hat{U}(\boldsymbol{\theta}_{e_i}) = \hat{E}_{p_i}\hat{U}(\boldsymbol{\theta}_g)$.
    
    \item \textbf{Iterative phase estimation algorithm (IPEA)}: Subsequently, the IPEA is performed using $\hat{U}(\boldsymbol{\theta}_g)$ for the ground state and $\{\hat{U}(\boldsymbol{\theta}_{e_i})\}$ for the excited ones in the state preparation. The eigenstate can be obtained with further projection of each state onto the closest eigenstate and the energy estimate can be refined.
\end{enumerate}
The detailed structure of the algorithm is illustrated in Fig.~\ref{fig_6}.

As a proof-of-concept, WAVES was applied to estimate the spectrum of a simplified exciton transfer Hamiltonian modeling two chlorophyll units in the LHII complex:
\begin{align}
    \hat{H} = \alpha\hat{I} + \beta \hat{X},  
\end{align}
where $\alpha=1.46~\mathrm{eV}$, $\beta =0.037~\mathrm{eV}$, and $\ell$ is a chosen energy shift to enhance IPEA resolution. The ground and excited states were found to have high fidelities up to 99\%, and IPEA was used to extract the eigenvalues up to 32 bits of precision. In summary, the WAVES algorithm combines the robustness of variational methods with the precision of phase estimation, offering a powerful approach for accurate spectrum estimation on near-term photonic quantum devices.

\begin{figure*}[t]
\centerline{\includegraphics[width=15cm]{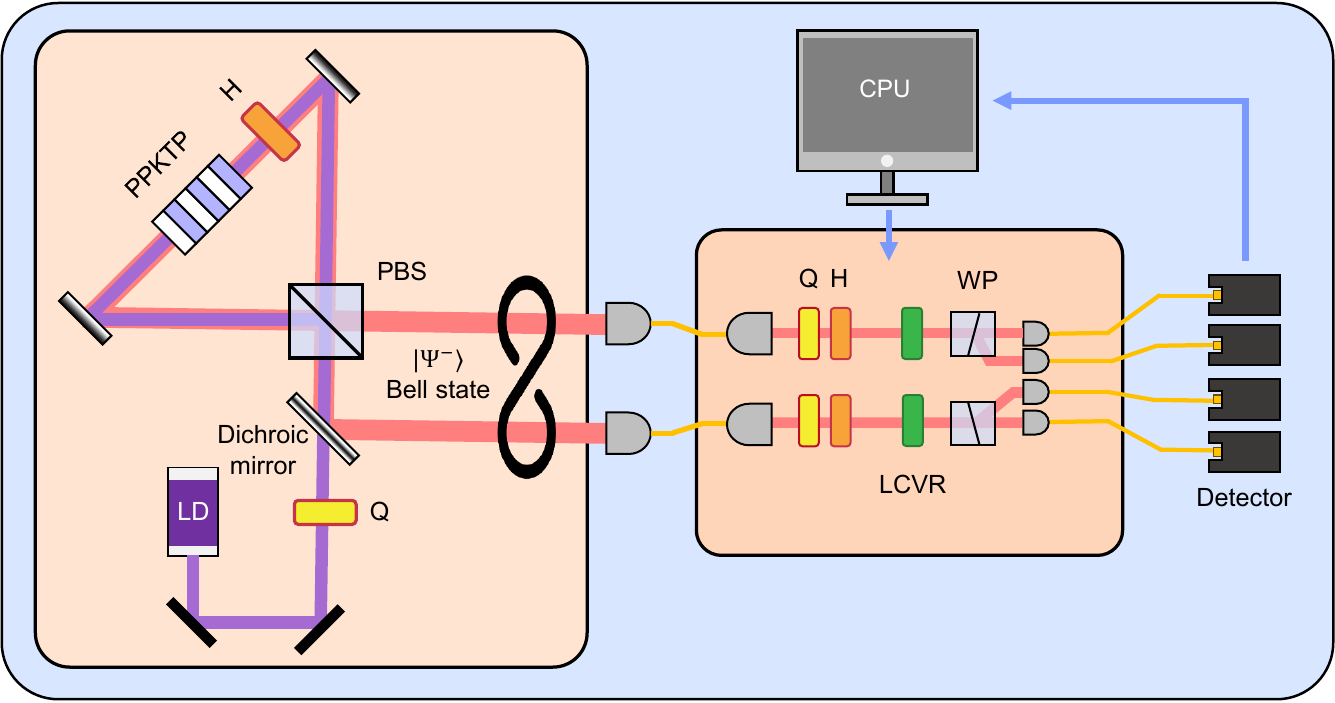}}
\caption{Experimental setup for VQE in Ref.~\cite{borzenkova2024}. An ansatz state can be controlled by Q and H. The dephasing noise can be implemented by LCVRs. (LD, laser diode; PPKTP, periodically poled potassium titanyl phosphate; PBS, polarizing beam-splitter; Q, quarter-wave plate; H, half-wave plate; LCVR, liquid crystal variable retarder; WP, Wollaston prism)}
\centering
\label{fig_7}
\end{figure*}

\subsubsection{VQE for physical models}
Since the VQE is, in principle, applicable to general eigenvalue problems, it has been employed not only in quantum chemistry but also in solving important physical models, such as the Heisenberg and Schwinger models.

For example, Ref.~\cite{borzenkova2021} demonstrated a photonic VQE implementation for solving the two-qubit Schwinger model, described by the Hamiltonian:
\begin{equation}
    \hat{H}(m)=\hat{I}\otimes\hat{I}+\hat{X}\otimes\hat{X}+\hat{Y}\otimes\hat{Y}-\frac{1}{2}\hat{Z}\otimes\hat{I}+\frac{1}{2}\hat{Z}\otimes\hat{Z}+\frac{m}{2}(\hat{I}\otimes\hat{Z}-\hat{Z}\otimes\hat{I}),
\end{equation}
where $m$ is the particle mass. This Hamiltonian has four non-degenerate eigenvalues $E_1,E_2,E_3,E_4$, among which $E_2=2$ and $E_3=1$ remain constant for any $m$. The other two eigenvalues, $E_1$ and $E_4$, depend on the mass and are given by:
\begin{eqnarray}
   E_{j}=\frac{1}{2}+(-1)^{j}\sqrt{m^2+m+\frac{17}{4}}, \ \ j=1,4.
\end{eqnarray}
In the experiment illustrated in Fig.~\ref{fig_7}, a Bell state $|\Psi^-\rangle$ was generated via SPDC process in a Sagnac interferometer. This state underwent unitary operators and Pauli measurements implemented using half-wave plates (HWPs) quarter-wave plates (QWPs), and Wollaston prisms, respectively. To test the VQE's noise tolerance, dephasing noise was introduced using a liquid crystal variable retarder. The results showed that the ground state energy could be reliably estimated under moderate dephasing, and that even under strong noise, the phase transition point at $m=-\frac{1}{2}$--an important feature of the Schwinger model--was still observable. These findings highlight the inherent robustness of VQE against certain types of noise.

\begin{figure}[t]
\centerline{\includegraphics[width=12cm]{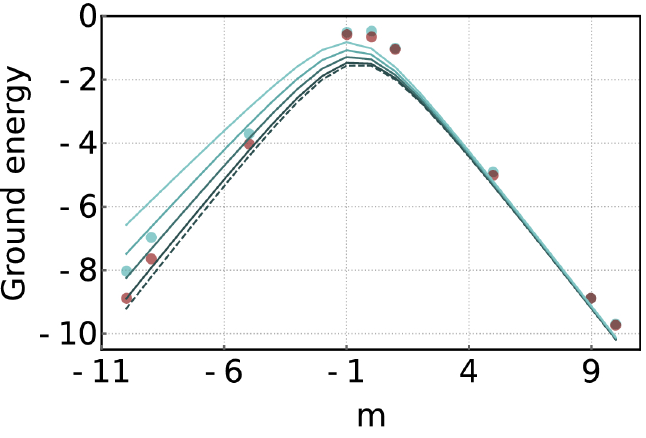}}
\caption{Experimental results from Ref.~\cite{borzenkova2024}. Four cyan lines represent simulated results for different noise strengths $\varepsilon_1$ = 0.1, 0.3, 0.5, 0.7. Cyan and red dots represent the experimental results without and with error-mitigation, respectively {(Reprinted with permission from Ref.~\cite{borzenkova2024} © Optical Society of America).}}
\centering
\label{fig_8}
\end{figure}

In a follow-up study~\cite{borzenkova2024}, ZNE was applied to further improve the accuracy of the VQE applied to the Schwinger model. To implement ZNE, the researchers experimentally characterized two intrinsic noise levels, $\varepsilon_1$ and $\varepsilon_2$, using the Hong-Ou-Mandel (HOM) effect~\cite{hong1987}. Specifically, HOM dip visibility $V$ of the generated two single-photons passing a fiber beam-splitter was measured, and the corresponding noise strength was calculated as $\varepsilon=1-V$. {For each set of variational parameters, the expectation value of the Hamiltonian was then evaluated at both noise levels, yielding $E(\epsilon_1)$ and $E(\epsilon_2)$, and assuming a linear dependence on $\epsilon$ in the low-noise regime, these two points were used to extrapolate to the zero-noise limit $E(0)$, which served as the effective cost function for the VQE optimization.}

Finally, the error-mitigated expectation value is obtained through the ZNE as illustrated in Fig.~\ref{fig_8}, and thus the indistinguishability-related noise occurring at the two-photon generation process is mitigated. In experiment, a pair of photons was generated via SPDC process and both HWP and QWP were used to estimate the intrinsic noise strengths. For example, the HWP is rotated by $10^{\circ}$ to consider the case of $\varepsilon_1=0.18$ and $\varepsilon_2=0.29$. The VQE was performed on a {photonic integrated circuit} with structure of Fig.~\ref{fig_5}(a). The experimental results are shown in Fig.~\ref{fig_8}~\cite{borzenkova2024}. The method was successfully demonstrated to mitigate experimental errors arising from differences in the properties of two-photons.

\subsubsection{VQE for factorization}
In addition to applications in quantum chemistry and physics, VQE has also been explored as a tool for integer factorization--a problem of fundamental importance in quantum computing~\cite{anschuetz2018, karamlou2021,agresti2024}. In Ref.~\cite{agresti2024}, the feasibility of performing factorization using VQE was experimentally demonstrated using a photonic integrated circuit. Specifically, the factorization of $N = 35 = 5 \times 7$ was encoded into the ground state of a specially constructed Ising-type Hamiltonian, and the energy landscape was optimized using a hardware-efficient ansatz (HEA) implemented on the photonic platform.

The factorization problem was first classically mapped to the minimization of the objective function $f(x, y) = (N - xy)^2$, whose minimum corresponds to the desired prime factors $p$ and $q$. For the specific case of $N = 35$, the $x$ and $y$ can be represented in binary number and the corresponding Hamiltonian is written as:
\begin{align}
    \hat{H} = \left[ N\hat{I}- \left(4\hat{I}+ 2\hat{\Pi}^{1}_z + \hat{I}\right)_1 \otimes \left(4\hat{I} + 2\hat{\Pi}^{1}_z + \hat{I}\right)_2 \right]^2,
\end{align}
where $\hat{\Pi}^{1}_z=\frac{\hat{I}-\hat{Z}}{2}$. The VQE is implemented on the {photonic integrated circuit}. The ground state energy was estimated using measurements in the computational basis only, leveraging the fact that the ground state is an incoherent mixture of $\ket{01}$ and $\ket{10}$, thus obviating the need for full state tomography. The result of VQE converged to one of the two degenerate ground states, corresponding to the correct binary-encoded solutions $x = 101 = 5$ and $y = 111 = 7$, or vice versa. These results demonstrate the potential of photonic systems for implementing variational protocols for non-quantum-chemical problems. The use of hardware-tailored ansatz circuits and efficient encoding strategies proves critical in extending the applicability of VQE to computational tasks such as integer factorization. 


\begin{figure*}[b]
\centerline{\includegraphics[width=16cm]{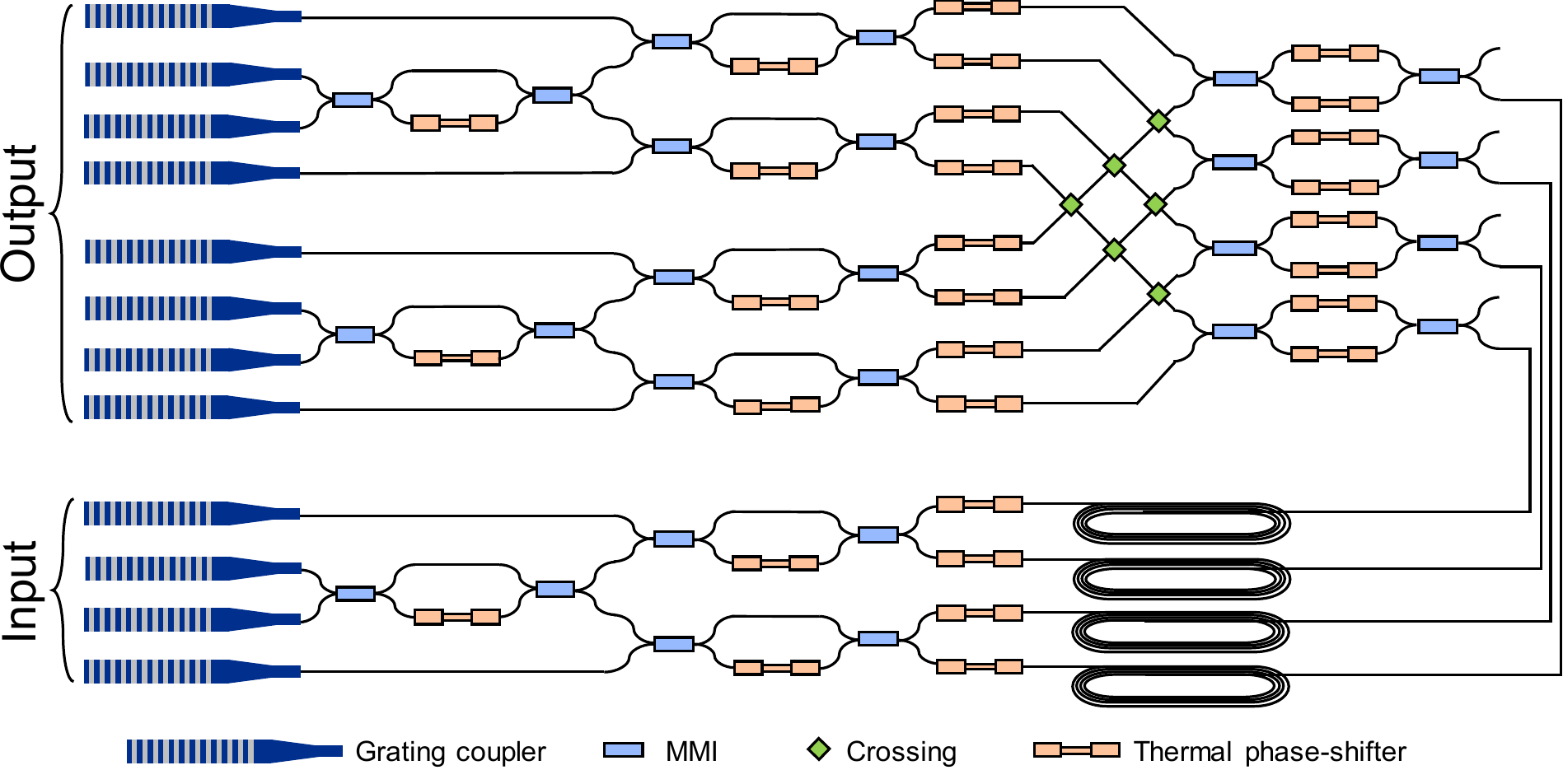}}
\caption{Experimental setup for factoring $35=5\times7$ using the VQE~\cite{baldazzi2025}. Reck's configuration~\cite{reck1994} was used to realize arbitrary mode transformation that splits input photons, including multimode-interference-based integrated beam splitters (MMI).}
\centering
\label{fig_9}
\end{figure*}

\subsection{{Scalability strategies}}
\subsubsection{Multi-qubit approach}
To solve more complex Hamiltonians, more than two qubits are typically required to capture the increased Hilbert space dimensionality. Several studies have demonstrated the scalability of photonic VQE systems toward this direction. 

In Ref.~\cite{ghavami2025}, a simulation-based design of a photonic integrated circuit was proposed for estimating the ground state energy of the He molecule, which required four qubits. A quantum circuit for preparing the ansatz was constructed to mimic a Mach-Zehnder interferometer (MZI) architecture, consisting of directional couplers and phase shifters. While the VQE was implemented using Qiskit simulation, this work demonstrated the scalability and architectural feasibility of photonic VQE operating in a $4$-dimensional Hilbert space.

More impressively, Ref.~\cite{baldazzi2025} reported an experimental realization of a $16$-dimensional VQE using a programmable photonic integrated circuit based on Reck's triangular interferometer design~\cite{reck1994} (See Fig.~\ref{fig_9}). This chip integrated $120$ thermo-optic phase shifters distributed across a mesh of 32 tunable beam-splitters, enabling the implementation of arbitrary $8$-mode unitary transformations on dual-rail encoded ququarts.

This versatile photonic platform was used to solve two distinct problems: the integer factorization of $35 = 5 \times 7$ and the ground state energy estimation of $\mathrm{H}_2$. For the integer factorization, a diagonal Ising Hamiltonian was encoded such that its ground state represented binary-encoded valid factor pairs. The VQE converged to one of the correct solutions, yielding either $\ket{0101}$ and $\ket{0111}$, corresponding to the factors $5$ and $7$ (or vice versa). On the other hand, for the ground state energy estimation of $\mathrm{H}_2$, a UCCSD ansatz was constructed, consisting of both local and entangling gates compiled into a single global unitary. This structure allowed the efficient exploration of the chemically relevant subspace within the large Hilbert space.

 In both cases, the SPSA algorithm was employed for classical optimization. The experimental results demonstrated high-fidelity agreement with theoretical expectations. Notably, the same hardware was used for both tasks without any physical modification, showcasing the reprogrammability of the platform. These results highlight the scalability, versatility, and practical controllability of photonic integrated circuits for VQE. By exploiting programmable architectures and robust optimization techniques, photonic systems are shown to be viable candidates for addressing increasingly complex quantum problems in the NISQ era.

\begin{figure*}[b]
\centerline{\includegraphics[width=12cm]{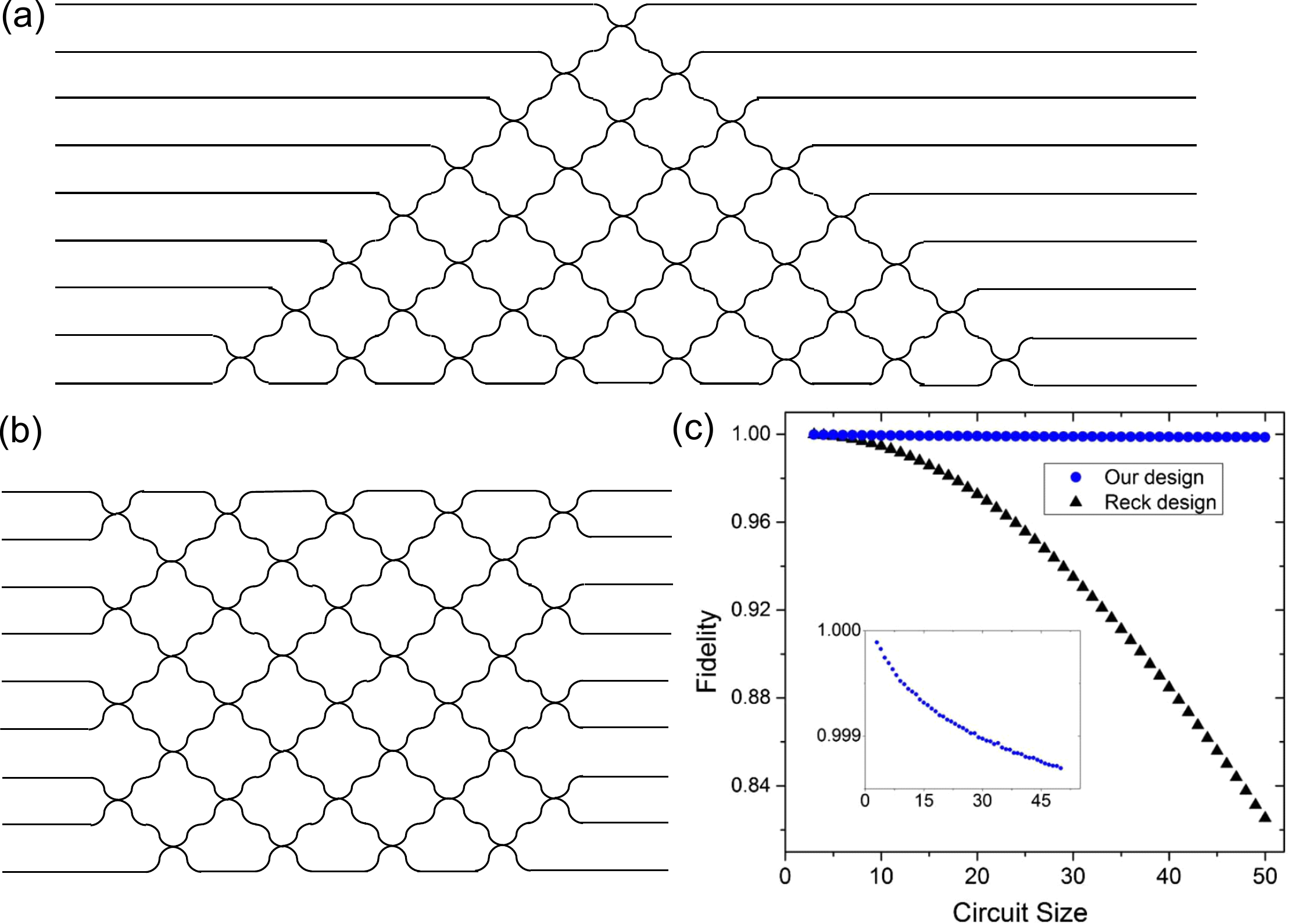}}
\caption{Configurations for implementing arbitrary mode transformations in a controllable beam-splitter network. (a) Reck's configuration~\cite{reck1994}, (b) Clements' configuration~\cite{clements2016}. (c) Comparison of fidelity under photon loss for beam-splitter networks, where Clements' configuration demonstrates superior robustness over Reck's configuration~\cite{clements2016} (Reprinted with permission from Ref.~\cite{clements2016} © Optical Society of America).}
\centering
\label{fig_10}
\end{figure*}

\begin{figure*}[b]
\centerline{\includegraphics[width=16cm]{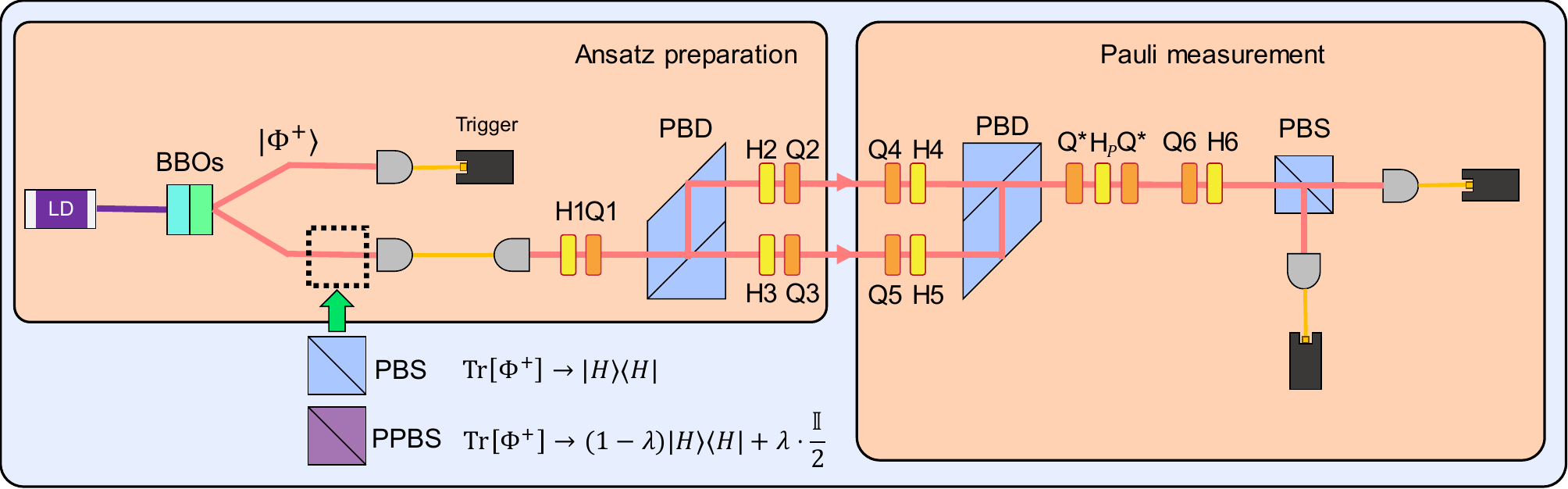}}
\caption{Experimental setup for the VQE implementation in Ref.~\cite{lee2022}. A laser diode pumps a pair of BBO nonlinear crystals arranged in a sandwich configuration, generating a two-photon entangled state $|\Phi^+\rangle$ via SPDC process. One of the photons is detected to herald a singe photon, which is then sent into the VQE setup. The input polarization state is filtered using either a PBS or a PPBS, with the latter introducing depolarizing noise. $\textrm{Q}^*$ is fixed at 45\textdegree \ and $\textrm{H}_p$ compensates for the phase drift between spatial modes (LD, laser diode; BBO, $\beta$-barium borate; PBS, polarizing beam-splitter; PPBS, partially polarizing beam-splitter; Q, quarter-wave plate; H, half-wave plate; PBD, polarizing beam displacer).}
\centering
\label{fig_11}
\end{figure*}

To improve scalability in photonic VQE, Clements' square interferometer design was proposed in Ref.~\cite{clements2016} as a photon-loss-tolerant architecture for constructing programmable beam-splitter networks. As shown in Fig.~\ref{fig_10}, this configuration arranges beam splitters in a square mesh with reduced circuit depth compared to Reck's triangular design, thereby minimizing cumulative loss. 

In this configuration, a beam-splitter network is composed of multiple beam-splitters described by the mode transformation
\begin{eqnarray}
    T_{m,n}=\begin{bmatrix}
       e^{i\phi_{m,n}}\cos\theta_{m,n} & -\sin\theta_{m,n} & \\
       \sin\theta_{m,n} & e^{i\phi_{m,n}}\cos\theta_{m,n} 
    \end{bmatrix},
\end{eqnarray}
where $T_{m,n}$ denotes the transformation between spatial modes $m$ and $n$. An arbitrary unitary transformation $U$ on the interferometer can then be decomposed as:
\begin{eqnarray}
    U=D\left(\prod_{(m,n)\in S}T_{m,n}\right),
\end{eqnarray}
where $D$ is a diagonal matrix composed of redundant phase terms such as $e^{i\delta_1},e^{i\delta_2},\cdots$. Notably, the Clements' configuration is robust to photon loss compared with the Reck's configuration, since it has shallow circuit depth. This was numerically demonstrated as shown in Fig.~\ref{fig_10}(c)~\cite{clements2016}.

\subsubsection{Single-qudit approach with error-mitigation}
Another promising approach toward scalability is the use of qudits, which harness the high-dimensional degrees of freedom of a single-photon. Since a single $d$-level qudit can encode the same amount of information as $\log_2d$ qubits, this strategy allows high-density quantum computation with fewer particles.

In Ref.~\cite{lee2022}, a $4$-dimensional photonic VQE was implemented using a qudit encoded by combining polarization and path degrees of freedom of a single-photon. The experimental setup is shown in Fig.~\ref{fig_11}. The qudit is initialized by generating a heralded Bell state $|\Phi^+\rangle$ via SPDC in a pair of $\beta$-barium borate (BBO) crystals. One single-photon is detected as a trigger, while the other one is directed into a qudit-encoding interferometer. The qudit ansatz is implemented by manipulating the single-photon with HWPs and QWPs in an MZI configuration. Expectation values of Pauli strings are subsequently obtained via measuring the polarization and path degrees of freedom of the single-photon. Based on the measurement outcome, parameters of the MZI configuration are iteratively updated using COBYLA, which was reported to outperform the other classical optimizers in this work. Here, depolarizing noise was experimentally realized using partially polarizing beam-splitter (PPBS) to demonstrate noise-tolerance of the method. The noise strength of the depolarizing noise was characterized to recover the expectation value using the Pauli noise error-mitigation. {Experimentally, the noise channel for each Pauli measurement is determined by sending its eigenstates through the full optical setup and using the resulting input–output statistics to reconstruct a classical transition matrix. The inverse of transition matrix is then applied in post-processing to correct the measured probability distributions.}

\begin{figure}[t]
\centerline{\includegraphics[width=12cm]{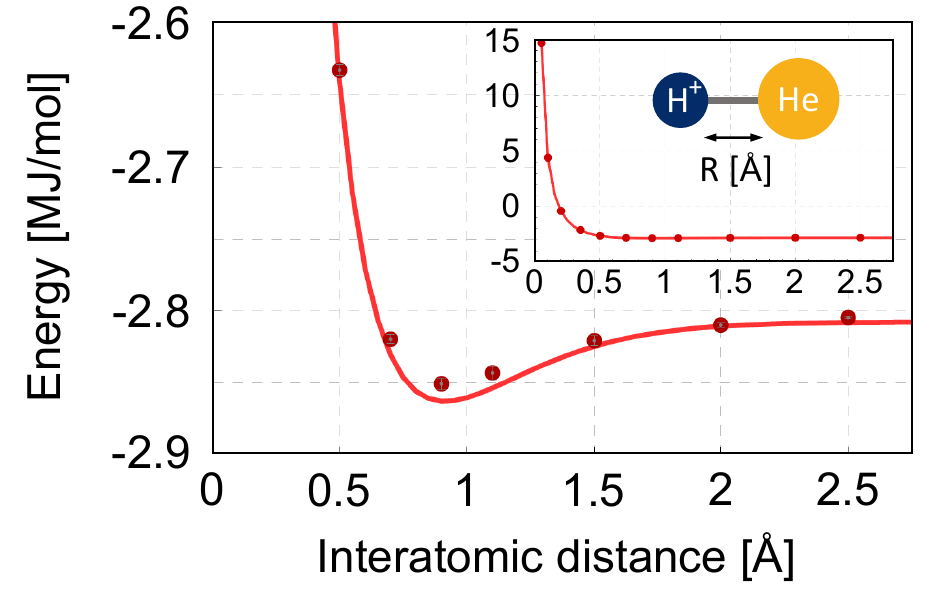}}
\caption{Experimental results from Ref.~\cite{lee2022}. The solid red curve represents the theoretical prediction, while the data points indicate the experimentally obtained values after Pauli noise error-mitigation (Reprinted with permission from Ref.~\cite{lee2022} © Optical Society of America).}
\centering
\label{fig_12}
\end{figure}

The final error-mitigated results, shown in Fig.~\ref{fig_12}~\cite{lee2022}, were in strong agreement with the theoretical ground state energy of the $\mathrm{HeH^+}$ Hamiltonian. This work marks the first experimental demonstration of Pauli noise error mitigation in photonic VQE, showing the feasibility of high-dimensional, noise-tolerant VQE with single-photon qudit encodings.

\subsubsection{VQE with quantum natural gradient optimizer}
In Ref.~\cite{wang2023}, a key focus was placed on the classical optimization routine in variational quantum algorithms (VQAs), recognizing that their overall performance crucially depends on the quality of the optimizer. The authors experimentally demonstrated the quantum natural gradient (QNG) method applied to VQE using a qudit-based photonic system. 
 
 Unlike conventional gradient-free optimizers such as Nelder–Mead or COBYLA, the QNG method incorporates the underlying geometry of the Hilbert space by rescaling the gradient with the inverse of the quantum Fisher information matrix (QFIM)~\cite{liu2020} 
\begin{eqnarray}
    F_{jk}(\boldsymbol{\theta})=4\mathrm{Re}\left\{\langle\partial_{\theta_j}\psi(\boldsymbol{\theta})|\partial_{\theta_k}\psi(\boldsymbol{\theta})\rangle-\langle\partial_{\theta_j}\psi(\boldsymbol{\theta})|\psi(\boldsymbol{\theta})\rangle\langle\psi(\boldsymbol{\theta})|\partial_{\theta_k}\psi(\boldsymbol{\theta})\rangle\right\},
\end{eqnarray}
where $|\partial_{\theta_j}\psi(\boldsymbol{\theta})\rangle$ denotes the partial derivative of $|\psi(\boldsymbol{\theta})\rangle$ with respect to the parameter $\theta_j$. Specifically, the iteration rule is defined to $\boldsymbol{\theta}_n=\boldsymbol{\theta}_{n-1}-\eta F^{-1}(\boldsymbol{\theta}_n)\vec{\nabla}E(\boldsymbol{\theta}_{n-1})$ where $F$ denotes the QFIM. Noting that the evaluation of QFIM requires expensive computational cost~\cite{yamamoto2019,stokes2020}, the SPSA is associated to efficiently evaluate the QFIM in the QNG method. Consequently, the QNG method used in this work, described by the iteration rule $\boldsymbol{\theta}_n=\boldsymbol{\theta}_{n-1}-\eta \bar{F}^{-\alpha}(\boldsymbol{\theta}_n)\vec{\nabla}E(\boldsymbol{\theta}_{n-1})$ with the approximated QFIM $\bar{F}$ and a real parameter $\alpha$, guarantees fast convergence speed, reduced fluctuation, and noise-tolerance. 

\begin{figure}[t]
\centerline{\includegraphics[width=10cm]{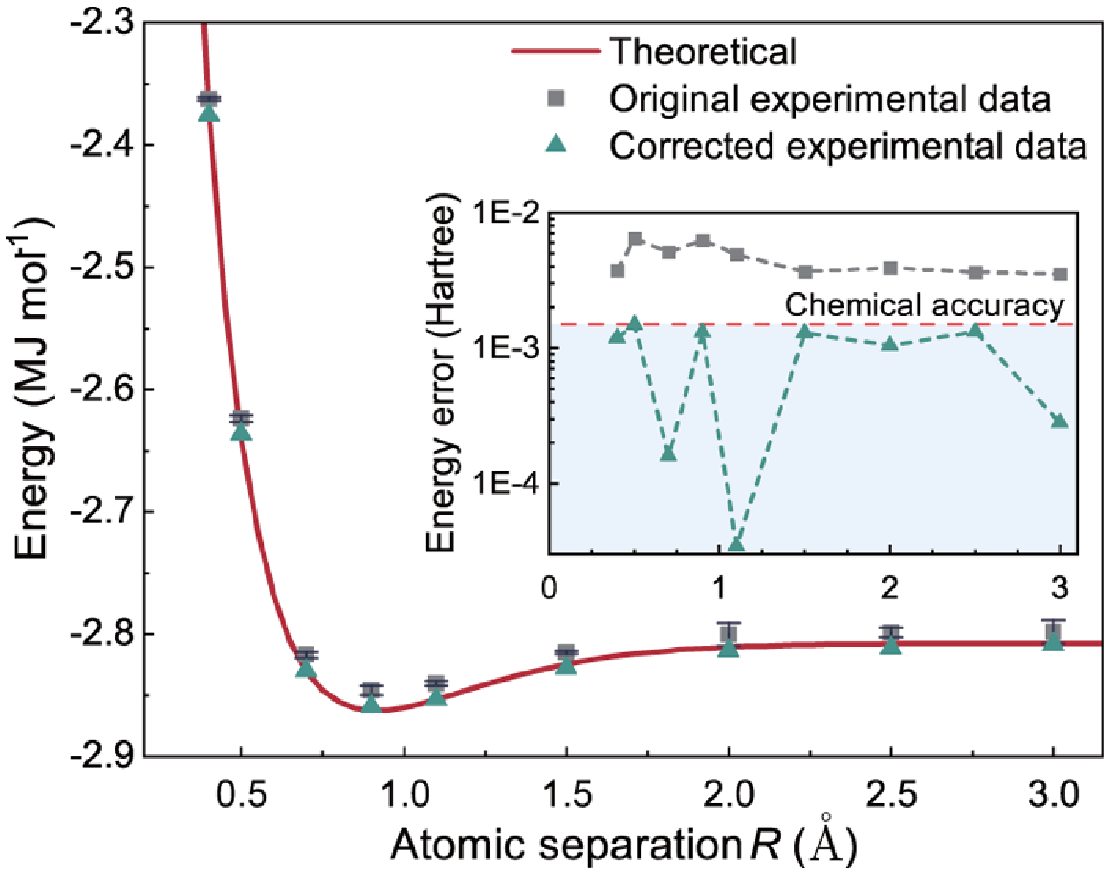}}
\caption{Experimental results from Ref.~\cite{wang2023}. The energy values and their associated error bars represent one standard deviation, obtained from 10 repetitions. The final energy for each run is calculated by averaging the measured energy values over the last five steps following convergence (Reprinted with permission from Ref.~\cite{wang2023} © Optical Society of America).} 
\centering
\label{fig_13}
\end{figure}

The experimental implementation was performed on a photonic integrated circuit using path-encoded qudit states to estimate the ground state energy of the ${\rm HeH^+}$ Hamiltonian. The variational ansatz was realized through a reconfigurable interferometric network, and expectation values were obtained by measuring the photon's path degree of freedom. Crucially, QNG updates were computed directly from experimental data, without relying on full state tomography. The results, presented in Fig.~\ref{fig_13}~\cite{wang2023}, demonstrate that the QNG method is both practically implementable and effective in the photonic VQE setting. This work represents the first experimental realization of photonic VQE incorporating QNG optimization, marking a notable advancement in developing NISQ-compatible quantum-classical hybrid algorithms.


\begin{figure*}[b]
\centerline{\includegraphics[width=15cm]{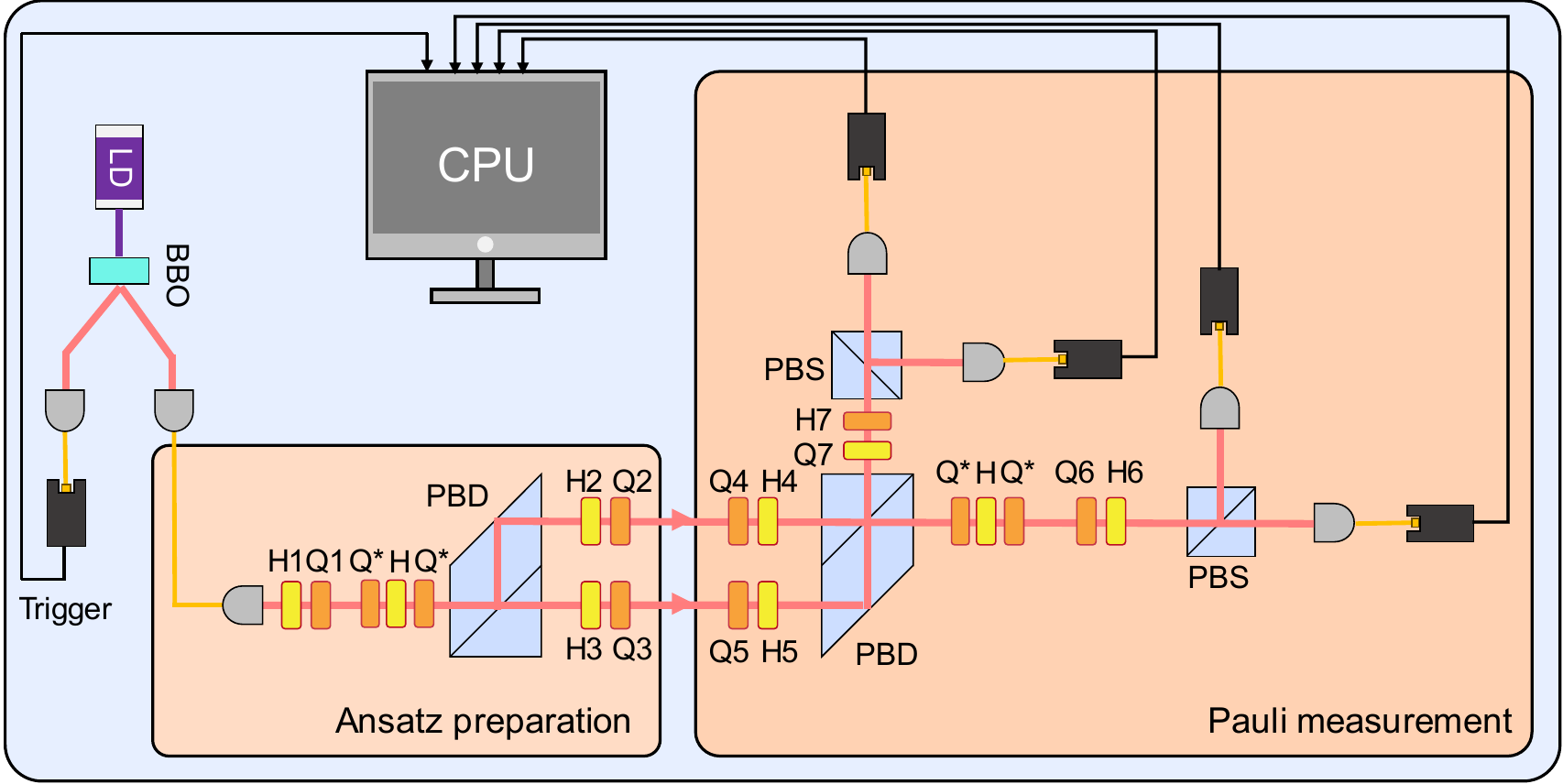}}
\caption{Experimental setup for VQE in Ref.~\cite{lee2024}. A 390 nm pulsed laser pumps a BBO nonlinear crystal, generating a pair of photons via SPDC. he idler photon is detected as a trigger, while the signal photon is encoded in both polarization and path degrees of freedom. Bell state measurements are implemented using a PBD and a PBS. $\textrm{Q}^*$ is fixed at 45\textdegree \ and $\textrm{H}_p$ compensate for the phase drift between spatial modes (LD, laser diode; BBO, $\beta$-barium borate; PBS, polarizing beam-splitter; Q, quarter-wave plate; H, half-wave plate; PBD, polarizing beam displacer).}
\centering
\label{fig_14}
\end{figure*}

\subsubsection{VQE using Bell measurement}
To reduce the number of measurements required in VQE, Ref.~\cite{lee2024} proposed using Bell measurements, which allow for simultaneous measurement of multiple Pauli strings. By grouping Pauli operators based on general commutativity (GC)--a less restrictive condition than qubit-wise commutativity (QWC)--the number of distinct measurement settings needed for full Hamiltonian evaluation can be significantly reduced.

For example, consider the case of the ${\rm HeH^+}$ Hamiltonian, which contains nine Pauli strings. A QWC-based grouping requires four measurement groups: $\{\hat{I}\otimes\hat{X}, \hat{Z}\otimes\hat{X}\}, \{\hat{X}\otimes\hat{I}, \hat{X}\otimes\hat{Z}\}, \{\hat{X}\otimes\hat{X}\},$ and $\{\hat{I}\otimes\hat{I}, \hat{I}\otimes\hat{Z}, \hat{Z}\otimes\hat{I}, \hat{Z}\otimes\hat{Z}\}$.

In contrast, GC-based grouping combined with Bell measurements reduces this to just three: $\{\hat{X}\otimes\hat{X}, \hat{Z}\otimes\hat{Z}, \hat{I}\otimes\hat{I}\}$, $\{\hat{X}\otimes\hat{I}, \hat{X}\otimes\hat{Z}, \hat{I}\otimes\hat{Z}\}$, and $\{\hat{I}\otimes\hat{X}, \hat{Z}\otimes\hat{I}, \hat{Z}\otimes\hat{X}\}$, since the first group is jointly measurable using the Bell measurement. The latter measurement grouping also lower the number of controlling wave plate configuration in the setup, thereby reducing the experimental cost of VQE. 

In particular, this method is highly effective for estimating the ground state energy of the two-qubit Schwinger model: $\hat{H}_{\rm 2q}=\hat{X}\otimes\hat{X}+\hat{Y}\otimes\hat{Y}+\hat{Z}\otimes\hat{Z}$. These Pauli strings can be decomposed into projectors onto the Bell basis:
\begin{eqnarray}
    \hat{X}\otimes\hat{X}&=&|\Psi^+\rangle\langle\Psi^+|+|\Phi^+\rangle\langle\Phi^+|-|\Psi^-\rangle\langle\Psi^-|-|\Phi^-\rangle\langle\Phi^-|,\nonumber\\
    \hat{Y}\otimes\hat{Y}&=&|\Psi^+\rangle\langle\Psi^+|+|\Phi^-\rangle\langle\Phi^-|-|\Psi^-\rangle\langle\Psi^-|-|\Phi^+\rangle\langle\Phi^+|,\nonumber\\
    \hat{Z}\otimes\hat{Z}&=&|\Phi^+\rangle\langle\Phi^+|+|\Phi^-\rangle\langle\Phi^-|-|\Psi^+\rangle\langle\Psi^+|-|\Psi^-\rangle\langle\Psi^-|.
\end{eqnarray}
Thus, a single Bell measurement can suffice to estimate all relevant expectation values, drastically reducing the required resources. 

As illustrated In Fig.~\ref{fig_14}, this strategy was implemented using a single-photon encoding scheme combining polarization and path degrees of freedom. A key feature of this encoding is that the Bell measurement can be realized using only linear optical elements. Specifically, a polarizing beam displacer (PBD) functions as a CNOT operation between the polarization (control) and path (target) qubits, enabling the Bell measurement that perfectly discriminates every Bell states. This HEA-based method facilitates the practical implementation of GC-based measurement grouping in VQE. This method significantly reduces the number of optical reconfigurations in the experiment, as each measurement group corresponds to a fixed wave plate configuration.

To evaluate its effectiveness, the authors experimentally estimated the ground state energy of the ${\rm HeH^+}$ molecule at various interatomic distances. The results using both Pauli and Bell measurements $(P+E)$ show slightly better agreement with the theoretical ground state energy than those using only Pauli measurements $(P)$, and most of the $(P+E)$ results were within the chemical accuracy threshold of 0.0016 Ha. These findings highlight that Bell-based measurement grouping not only improves measurement efficiency but also enhances accuracy, by reducing experimental errors stemming from reconfiguration or optical imperfections.

\subsubsection{VQE with orbital angular momentum qudit states}
Among the various degrees of freedom for preparing and manipulating a qudit state, the orbital angular momentum (OAM) of a single-photon is an attractive and scalable option~\cite{yao2011, rambach2021,chen2019}. OAM-based qudit states, which in principle span an infinite-dimensional Hilbert space, can be practically accessed via hundreds of Laguerre--Gaussian (LG) modes. Numerous studies have demonstrated the application of such OAM qudit states in high-dimensional quantum information processing~\cite{fickler2012,fickler2016,krenn2014}. 

\begin{figure}[b]
\centerline{\includegraphics[width=14cm]{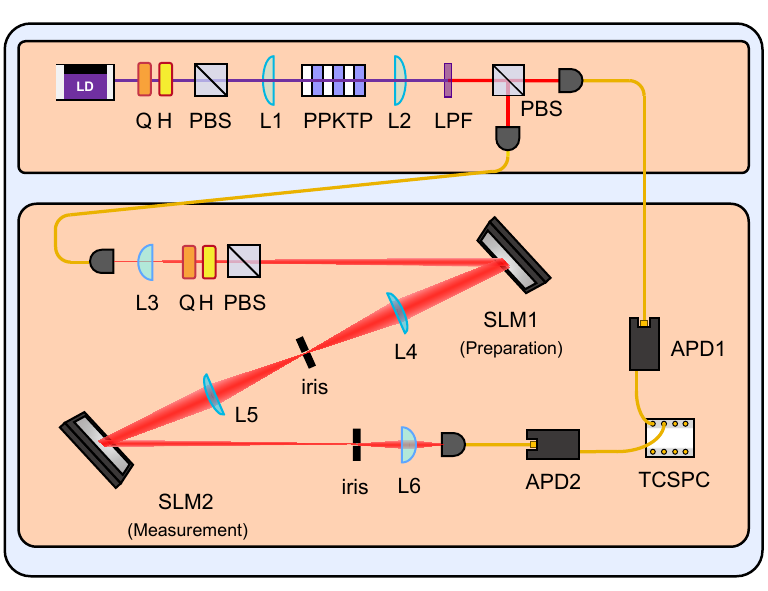}}
\caption{Experimental setup for VQE using OAM qudit states, as illustrated in Ref.~\cite{kim2024}. A heralded single-photon, generated via type-II SPDC process in a PPKTP crystal, is encoded into  an OAM mode using the SLM1. The prepared ansatz state is projected onto a Pauli basis using the SLM2, and then coupled into an optical single-mode fiber for detection (LD, laser diode; PPKTP, periodically poled potassium titanyl phosphate; PBS, polarizing beam-splitter; Q, quarter-wave plate; H, half-wave plate; L, lens; LPF, long pass filter; SLM, spatial light modulator; APD, avalanche photodiode; TCSPC, time-correlated single-photon counter).}
\centering
\label{fig_15}
\end{figure}

LG modes are the solutions of the paraxial Helmholtz equation in cylindrical coordinates, characterized by their OAM indices. In Ref.~\cite{kim2024}, a resource-efficient implementation of the VQE was demonstrated using single-photon OAM qudit states, as illustrated in Fig.~\ref{fig_15}. Compared to multi-qubit systems that require complex controlled gates to address high-dimensional problems, the OAM-based system achieves this using only holographic images, significantly reducing circuit complexity and physical resource requirements.

A general pure state in a $d$-dimensional Hilbert space is characterized by $2d-2$ real parameters. Notably, exploiting a single OAM degree of freedom allows one to prepare such states without additional constraints on the structure of the required unitary operators--the optical architecture remains independent of the qudit dimension, enhancing scalability.

\begin{figure}[b]
\centerline{\includegraphics[width=16cm]{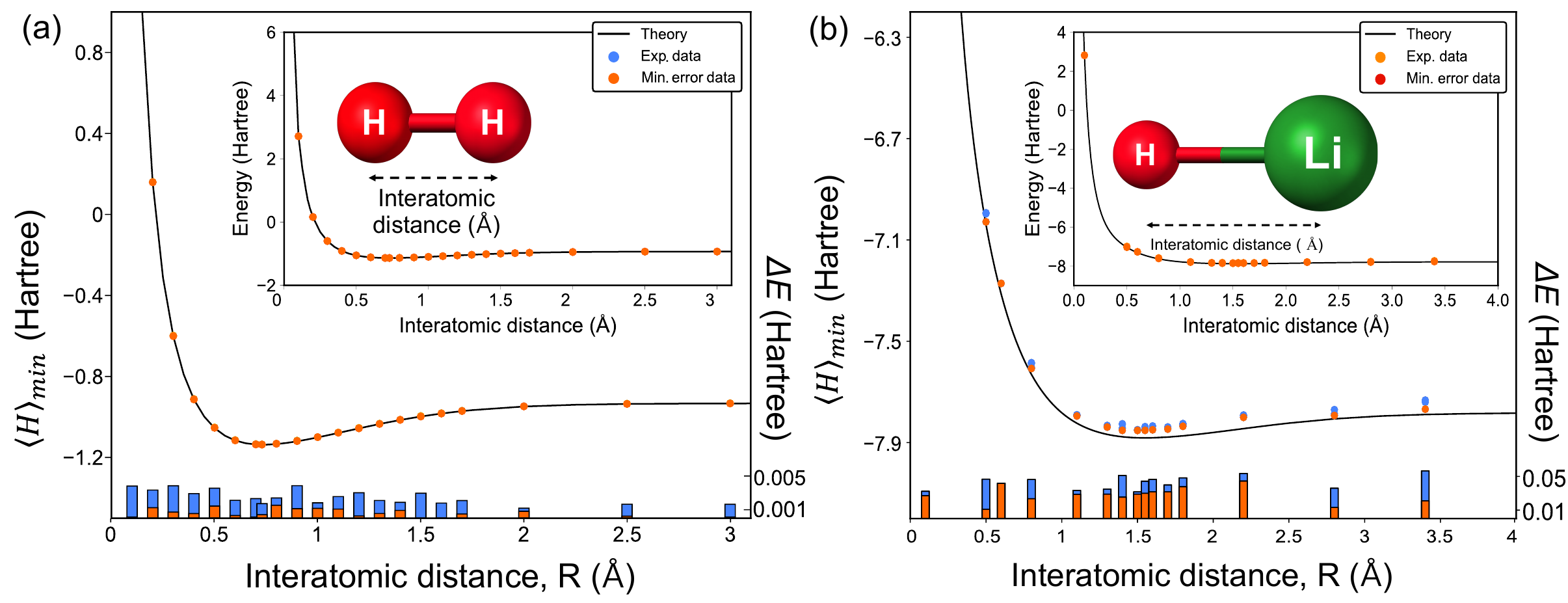}}
\caption{{(a) Experimental data from Ref.~\cite{kim2024} for estimating the ground state energy of $\mathrm{H}_2$ molecule. (b) Experimental data from Ref.~\cite{kim2024} for estimating the ground state energy of $\mathrm{LiH}$ molecule. In both (a) and (b), the black curve represent theoretical values, while the orange points indicate experimentally estimated ground state energy. The bars at the bottom show the differences between the theoretical values and the experimental results.}}
\centering
\label{fig_16}
\end{figure}

In this implementation, the VQE was demonstrated for both a $4$-dimensional ${\rm H_2}$ model and a 16-dimensional LiH model. The experimental setup illustrated in Fig.~\ref{fig_15} operates as follows: a heralded single-photon, which is generated via type-II SPDC process, propagates to a spatial light modulator (SLM1) with displaying a holographic image then the ansatz is prepared with superposition of LG modes. This qudit state is projected into a Pauli string basis by the measurement SLM (SLM2) and coupled into a single mode fiber. This scheme enables state preparation and measurement without controlled gates, reducing both circuit complexity and physical resources. 

For the ${\rm H_2}$ simulation, OAM indices $l\in\{-2,-1,1,2\}$  were used. As shown in Fig.~\ref{fig_16}(a), the VQE successfully achieved chemical accuracy without any error mitigation. For the more complex LiH molecule, indices $l\in\{-8,\cdots,-1,1,\cdots,8\}$ were considered. The VQE estimated the ground state energy with an average deviation from the theoretical value of approximately 0.036 $\pm$ 0.0098 Ha, as illustrated in Fig.~\ref{fig_16}(b). This work highlights the potential of single-photon OAM qudits for scalable quantum simulation, offering a compact, gate-free, and high-dimensional alternative to conventional multi-qubit VQE implementations.

\section{Conclusion}\label{sec4}

We have provided a comprehensive overview of the recent progress in the photonic implementation of variational quantum eigensolver (VQE), emphasizing both its theoretical underpinnings and experimental advancements. Photonic VQE has been successfully applied to tackle various problems in quantum chemistry, many-body physics, and integer factorization, even under the influence of noise. These demonstrations firmly establish photonic VQE as a promising candidate for NISQ-compatible quantum computing. The photonic VQE holds considerable potential for practical applications in various tasks such as {the design of new drug molecules~\cite{prakash2025} and the discovery of advanced battery material~\cite{hoang2024}, where quantum chemical simulations of molecular structure can provide useful insight alongside other experimental and engineering considerations.} 

Photonic systems offer several inherent advantages: they can generate, manipulate, and transmit qudit states with high fidelity at room temperature, and support high-dimensional encoding. These capabilities are particularly beneficial for scalable quantum networks and distributed quantum computing architectures~\cite{jeremy2007,jeremy2009,madsen2022,psiquantum2024}. When combined with programmable photonic integrated circuits, low-loss optical components, and advanced error-mitigation techniques, photonic VQE emerges as a compelling platform for practical quantum computation in the NISQ era.

Moreover, the VQE framework is not limited to estimating ground state energies. It can also be extended to compute excited state energies, which are crucial for understanding photochemical reactions, optical material properties, and energy transfer dynamics in complex systems. Several methods have been proposed for this purpose, including subspace-search VQE (SS-VQE)~\cite{nakanishi2019}, variational quantum deflation (VQD)~\cite{higgot2019}, multistate contracted VQE (MC-VQE)~\cite{parrish2019}, quantum subspace expansion (QSE)~\cite{colless2018}, constrained VQE~\cite{ilya2018}, and discriminative VQE~\cite{tilly2020}. Incorporating these methods into photonic VQE platforms will broaden their application to full-spectrum quantum spectroscopy and quantum dynamics simulations, thus narrowing the gap between near-term quantum processors and real-world impactful applications. This convergence of hardware scalability, error resilience, and algorithmic versatility positions photonic VQE as a pivotal component in the advancement of quantum information science and technology. {For the clarity, we also remark that the aforementioned VQE families are dependent on the numerical optimization. It implies that one needs to address their limitation such as the barren plateau problem~\cite{arrasmith2021}.}






\end{document}